\documentclass[twocolumn]{aastex631}
\usepackage{array}
\usepackage{amsmath}

\newcommand\Chandra{\textit{Chandra}}
\newcommand\SgrA{Sgr~A*}
\newcommand\ironK{Fe K$\alpha$}
\newcommand\solm{$M_\odot$}

\begin{document}

\title{Isolating Sgr A East: The First Uncontaminated X-ray Maps of a Galactic Center Supernova Remnant}

\correspondingauthor{Mayura Balakrishnan}
\email{bmayura@umich.edu}

\author[0000-0001-9641-6550]{Mayura Balakrishnan}
\affil{Trottier Space Institute at McGill University 
3550 Rue University, H3A 0C6, Montreal, QC, Canada}

\author[0000-0003-2251-6297]{Adrien Picquenot}
\affil{Department of Physics, Louisiana State University,  Baton Rouge, LA, USA}

\author[0000-0002-5466-3817]{Lia Corrales}
\affil{Department of Astronomy \& Astrophysics, University of Michigan, 311 West Hall, 1085 S. University \\
Ann Arbor, MI 48109, USA}

\author[0000-0002-9279-4041]{Q. Daniel Wang}
\affil{Department of Astronomy, University of Massachusetts Amherst, MA 21003, USA}

\author[0000-0002-6606-2816]{Fabio Acero}
\affil{FSLAC IRL 2009, CNRS/IAC, La Laguna, Tenerife}
\affil{Universit\'e Paris-Saclay, Universit\'e Paris Cit\'e, CEA, CNRS, AIM, F-91191 Gif-sur-Yvette Cedex, France}

\author[0000-0002-6752-2909]{Rodolfo Montez Jr.}
\affil{Center for Astrophysics $\vert$ Harvard \& Smithsonian, Cambridge, MA, USA}



\begin{abstract}

The central few parsecs of the Milky Way host a complex X-ray–emitting environment in which several extended plasma components are blended along the line of sight, complicating attempts to measure the intrinsic properties of individual components. In particular, the supernova remnant (SNR) Sgr A East is strongly confused with the stellar wind–fed plasma associated with Sagittarius A* and the surrounding nuclear environment. Here we apply Poissonian Generalized Morphological Component Analysis (pGMCA) to deep, stacked Chandra ACIS-I observations of the Galactic Center to disentangle these overlapping X-ray components. By comparing the separated X-ray components with multiwavelength data, we identify the location of the reflected shock in Sgr A East and construct spatially resolved maps of Fe and S/Ar/Ca emission. The Fe emission is centrally concentrated, consistent with the properties of mixed-morphology supernova remnants. Separating the SNR emission from the shocked wind plasma around \SgrA\ allows us to recover uncontaminated SNR properties and improve the robustness of the derived parameters. Spectral modeling of the isolated Sgr A East component reveals a lower ionization age and a higher electron density than previously reported, indicating strong interaction with dense surrounding material.
\end{abstract}



\section{Introduction} \label{sec:intro}

Supernova remnant (SNR) Sgr A East lives in one of the most intricate and densely populated regions in the Milky Way, with its iron emission peaking $\sim 1$ arcmin in projected distance from 4.5 $\times$ 10$^6$ \solm central supermassive black hole (SMBH) Sagittarius~A* (\SgrA). The plasma environment, which includes the supernova remnant (SNR) Sgr A East and wind-fed plasma surrounding \SgrA, as well as the accumulated stellar contribution from unresolved sources, overlaps along our line-of-sight, which complicates the analysis of each and renders previous efforts to distinguish the X-ray structures of the remnant from the surrounding emission incredibly challenging.

The remnant also appears to envelope the mini-spiral Sgr A West, a structure comprising three distinct streams of ionized gas channeling material from the circumnuclear disk (CND) to \SgrA. This reservoir of dense, warm molecular and atomic gas, with a mass of $10^5$–$10^6$ \solm, spans 1.5–3 parsecs from \SgrA. As infalling gas approaches the black hole, it becomes ionized by intense ultraviolet radiation from the nuclear star cluster (NSC), which hosts numerous young, massive stars \citep{Martins2007}. Many of these are Wolf-Rayet (WR) stars, whose powerful winds generate shocks that heat the surrounding medium to X-ray-emitting temperatures, creating the extended emission around \SgrA \citep{Cuadra2015, Calderon2020}. The NSC, containing $\sim 9.3 \times 10^6$ \solm\ within the central 4 parsecs \ \citep[100 arcseconds;][]{Chatzopoulos2015}, exerts a strong wind \citep{Rockefeller2005} that may even have influenced Sgr A East’s development \citep[e.g.,][]{Ehlerova2022}. Despite the extreme tidal forces and radiation environment near \SgrA, a small but dynamically important population of young ($\sim$4–6 Myr), massive stars is present, spatially and kinematically distinct from the old nuclear star cluster \citep{Paumard2006,Martins2007}, challenging conventional expectations for  star formation in such a hostile region. Investigating Sgr A East, whose progenitor may have been a member of the NSC, may provide insight into the formation of this young stellar population.

Sgr A East belongs to a rare class of supernova remnants known as mixed-morphology SNRs (MMSNRs). These remnants are characterized by thermal X-ray-filled centers surrounded by radio-bright shells, a morphology that deviates from standard SNR evolution models. MMSNRs, such as W44, W28, 3C391, and W49B, often exhibit associations with molecular clouds and 1720 MHz OH masers, indicating interactions with dense ambient media \citep[see][and references therein]{Vink2012}. The close association of these remnants with dense interstellar material underscores the importance of understanding their evolution, as \textit{their ejecta are most directly linked} to enriching the medium from which future generations of stars will form. The origin of the centrally peaked X-ray emission and the tendency for these SNRs to have signatures of overionization remains a topic of debate. One proposed explanation is thermal conduction, which redistributes heat and increases the central density while smoothing temperature gradients \citep{Cox1999}. Another scenario involves the evaporation of ambient cloudlets engulfed by the SNR shock, which increases the density in the remnant's interior \citep{WL1991}. Like many MMSNRs, Sgr A East is in close proximity and has demonstrated interaction with nearby molecular cloud M0.02-0.07 (also known as the +50 km/s cloud) through detection of masers \citep{1720MHzpaper}. Unlike most MMSNRs, however, the remnant is located near a significant energy source that may have supplied the additional energy required for overionization. Rather than resulting from interactions with dense ambient material, the overionization in Sgr A East may have been driven by past activity of \SgrA\ \citep{XRISMCollaboration2024}, potentially influencing the remnant’s evolution. 

Sgr A East displays a complex set of spectral and morphological properties that reflect its evolution in this unique and densely populated environment. In radio, it appears as a nonthermal shell \citep{YZ1986}, while in X-rays it has very concentrated 6.7 keV Fe K$\alpha$ emission, with an X-ray centroid offset by about 5 pc (or just over 2'; at the distance of \SgrA, 25"$\sim$1\ parsec) from the location of \SgrA \citep{Maeda2002,Park2004}. The X-ray core appears to extend only 2 arcmin wide \citep{Maeda2002}, in comparison to the radio shell which is $\sim$ 10' across \citep{YZ1987}, suggesting significant interaction with gaseous structures nearby. The presence of OH masers, which trace shocked gas, confirm that it has collided with and possibly compressed nearby molecular clouds \citep[MCs;]{YZ1999,YZ2010,Tsuboi2011}. Overlapping molecular cloud contours on X-ray emission, \citet{Park2004} found that Sgr A East is unlikely to be interacting with the 20 km/s cloud (M0.130.08), but has had its expansion inhibited by the 50 km/s cloud located northeast of \SgrA \citep{Tsuboi2011}. Absorption features from Sgr A West indicate the majority of the remnant is behind \SgrA \citep[]{YZ1987}. The concentrated iron core implies that a reverse shock has already propagated inward, reheating the ejecta \citep[]{Maeda2002,Park2004}. The morphology of the iron core further implies that the supernova explosion that created Sgr A East may have been asymmetric or heavily influenced by the gravitational potential of the Galactic Center, which could have shaped the distribution of ejecta and the dynamics of the shock waves \citep{Borkowski2013}. Chandra \citep{Ono2019}and XRISM \citep{XRISMCollaboration2024} spectral analysis of the remnant has indicated the presence of recombination lines, indicative of overionization that could be due to past \SgrA activity, or driven by other processes such as thermal conduction or adiabatic expansion \citep{YZ2010, Zhao2013}. Its evolution in this complex environment has complicated estimates of some significant properties, such as its age.

Estimates of the age of Sgr A East vary, but it is generally thought to be around 10,000 years old. For instance, if the gas in the region were moving at velocities typical of noncircular motions near the Galactic nucleus ($\sim$50 km/s), a displacement of 2.5 parsecs could be the result of a supernova event approximately 50,000 years ago \citep{YZ1987}. \citet{Maeda2002} estimated an age of 12.5 kyr using the Sedov-Taylor model, assuming a shock velocity of 1000 km/s (derived from iron line broadening) and an ambient density of 10$^3$ cm$^{-3}$, while accounting for shearing by Galactic rotation. On the other hand, \citet{Tsuboi2011} estimated an older age of 200 kyr, citing shock velocities around 50 km/s derived from CS/CO line measurements. However, this figure likely overstates the age as it does not factor in the turbulence present within the surrounding cloud structures. Simulations that incorporate a wind generated by the NSC generally suggest a younger age of 1,000 to 2,000 years \citep{Rockefeller2005, Yalinewich2017}.  The cannonball pulsar (CXO J174540.0-290031), moving at $\sim$500-1000 km/s away from the Sgr A East center, has been hypothesized to be the associated with the remnant, and if so dates the remnant to approximately 9000 years old. 

A deeper understanding of Sgr A East offers valuable insights into several astrophysical questions. For instance, how has the remnant's evolution been shaped by the wind generated by the nuclear star cluster? Was the progenitor of Sgr A East a Type Iax or Type II supernova, and what can the remnant's properties reveal about the progenitor star? The spatial distribution of iron-rich ejecta and the hot plasma cavity within Sgr A East can provide clues about the explosion dynamics and the nature of the progenitor. Additionally, the past expansion of Sgr A East, if accurately dated, could have triggered past outbursts of \SgrA \citep[see, e.g.,][]{baganoff2003}. Furthermore, studying Sgr A East as an MMSNR can shed light on the broader population of mixed-morphology remnants, particularly in terms of their interaction with dense environments and the mechanisms driving their unique X-ray morphologies. In this work, we aim to unravel the complex interactions between the various substructures within the central few arcminutes of the Galactic Center. Using 1.5 Ms of \Chandra\ ACIS-I observations, we apply signal separation technique Poissonian Generalized Morphological Component Analysis (pGMCA) to decompose the extended X-ray emission surrounding \SgrA. By isolating distinct substructures and analyzing their contributions, we aim to elucidate the dynamical processes and evolutionary history of the supernova remnant.

Section \ref{sec:data} outlines the \Chandra\ data reduction and stacking procedures, as well as the supplementary multiwavelength datasets incorporated to construct a comprehensive interpretation of the underlying mechanisms. In Section \ref{sec:data_exploration}, we conduct an exploratory analysis of the X-ray data, identifying salient features that inform subsequent pGMCA decomposition in Section \ref{sec:gmca_results}. Section \ref{sec:gmca_results} contextualizes the pGMCA-derived components within a multiwavelength framework to derive their physical properties. In Section \ref{sec:spectral_fits}, we use the results from pGMCA to conduct a spectral fit of the isolated SNR emission. Finally, we conclude with a synthesis of our findings.

\section{Datasets} \label{sec:data}

First, we describe the complementary multiwavelength datasets used to build a complete physical understanding of the X-ray plasma components isolated by signal separation technique pGMCA. Next, we detail the processing and stacking of all the archival Chandra ACIS-I observations used in this work to study the diffuse and extended X-ray plasma structures. The exact procedures taken to remove point sources and create diffuse X-ray images can be found in Appendix \ref{appendix_sec:stacking}. All analyses were conducted with CIAO 4.12, Heasoft 6.35.2, and CALDB 4.9.2.1.

\subsection{Multiwavelength Datasets} \label{sec:mw_data}

\begin{figure*}
    \centering
    \includegraphics[width=0.98\textwidth]{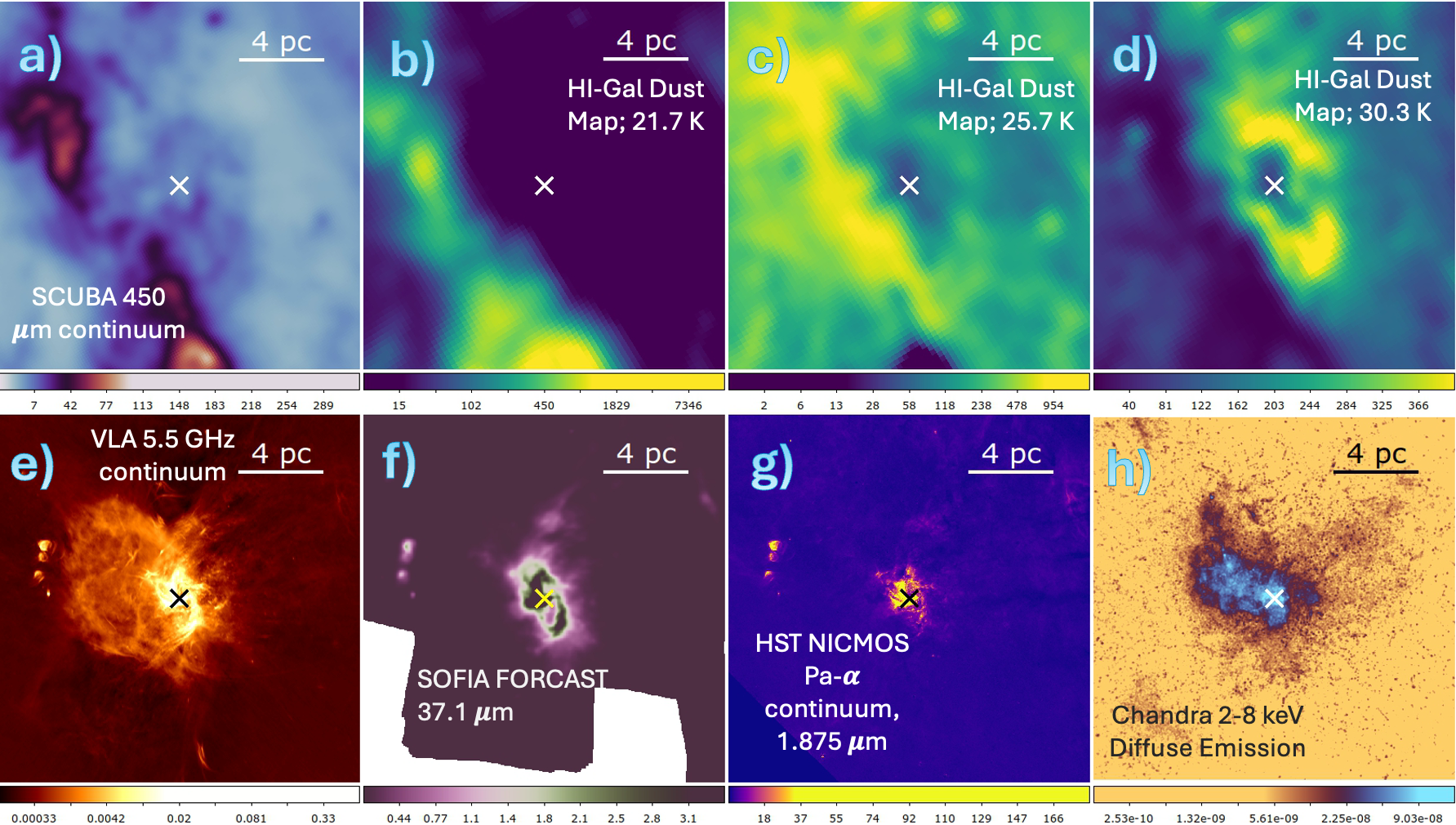}
    \caption{View of the region of interest in several different wavelength bands. In all panels, north is up and east is to the left, and an ``X" denotes the location of \SgrA. At the distance of \SgrA, 1 pc is approximately 25". \textit{a)} SCUBA 450\micron\ continuum image, showing the distribution and density of the molecular gas; \textit{b) - d)} Herschel HI-Gal dust temperature maps, reflecting the column density of material at a given dust temperature (labelled) along that line of sight \citep{Marsh2017} ; \textit{e)} VLA 5.5 GHz continuum image of the region, showing the extent of the supernova remnant shell \citep{Zhao2016}; \textit{f)} SOFIA FORCAST 37.1\micron\ emission, characterizing the warm dust in the region \citep{Lau2015}; \textit{g)} HST NICMOS 1.875 \micron\ Pa-$\alpha$ continuum \citep{Wang2010}; \textit{g)} Chandra 2-8 keV diffuse emission, with point sources removed, showing the major X-ray components along the line of sight: the wind-fed plasma surrounding \SgrA and the supernova remnant (this work).
. 
    }
    \label{fig:multiwavelength}
\end{figure*}

To get a better idea of the landscape surrounding Sgr A East, we used images and datasets from several published works and/or publicly accessible archival data. Some of the more visually useful datasets are pictured in Figure \ref{fig:multiwavelength}, while all are briefly described in bullet point form below. 

\begin{itemize}
    
    \item SCUBA (on the James Clark Maxwell Telescope) CMZ Survey \citep{Pierce-Price2000}: Traces dense molecular material. The image shows the molecular filament into which Sgr A East appears to be expanding, made up of the ``50 $\mathrm{km \, s^{-1}}$" and ``20 $\mathrm{km \, s^{-1}}$" clouds as they have been referred to previously.  (Figure \ref{fig:multiwavelength}; panel A)
    \item PPMAPS: Dust temperature maps from the Herschel HI-Gal survey \citep{Marsh2017}. The output units represent the column density of material with dust at a specific temperature for each sky position. For example, a 10 K map shows column densities (in $N_{H_2}$) where dust grains are near 10 K. The maps we use show the gas density associated with dust ranging from 20 to 50 K. The morphology of the dust changes greatly between 21 and 37 K, in particular.  (Figure \ref{fig:multiwavelength}; panels B-D)
    \item VLA 5.5 GHz continuum image \citep{Zhao2016}: Highlights the radio shell of the supernova remnant and the four HII regions to the east of Sgr A East. Also shows some radio cavities and some protruding emission to the west. (Figure \ref{fig:multiwavelength}; panel E)
    \item SOFIA FORCAST 37.1\micron\ warm dust emission (T$_{\rm dust} \sim$ 100 K) \citep{Lau2015}: The warm dust emission morphology is very different to any other dataset. The nearby HII regions east and northwest of Sgr A East are very bright. In addition, there is material that seems to surround the minispiral and co-spatial with plasma in the CND. There is also a protrusion in the west that is cospatial with the Western plume in the radio data. (Figure \ref{fig:multiwavelength}; panel F)
    \item Pa-$\alpha$ 1.875 \micron\ continuum from the HST NICMOS Galactic survey. Point sources have been removed and the image has been processed. The emission in this panel arises from hot H gas that is recombining (5000-15000 K) \citep{Wang2010}. (Figure \ref{fig:multiwavelength}; panel G)
    \item ALMA observations of G0.02-0.07 (the 50 $\mathrm{km \, s^{-1}}$ molecular cloud) \citep{Tsuboi2019}: We use SiO emission maps at a range of velocities. This gas is a tracer of shocked material, with concentrations in the minispiral and numerous discrete blobs. The emission seems to be clustered at $\sim$ 0 $\mathrm{km \, s^{-1}}$ and $\sim$ 50 $\mathrm{km \, s^{-1}}$, suggesting two major interaction fronts.
    \item 1720 MHz OH Masers \citep{Sjouwerman2008} and Class I 44 GHz Methanol Masers \citep{McEwen2016}, showing interaction sites at a variety of velocities.
\end{itemize}

\subsection{Chandra Data} \label{sec:ch_data}

We utilize 1.5 Ms of Chandra archival ACIS-I observations, spanning over two decades, listed in Table \ref{tab:chandra_data} and contained in~\dataset[DOI: 10.25574/cdc.559]{https://doi.org/10.25574/cdc.559}. We also investigated the possibility of utilizing ACIS-S observations, as there are 3 Ms of ACIS-S exposures taken of \SgrA\ and the instrument provides higher spectral resolution than ACIS-I. However, most of these observations were taken in a configuration that excludes a lot of the supernova remnant. All observations during which the Galactic Center transient AX J1745 (located southwest of \SgrA) was in outburst were excluded from our analysis. Due to its extreme brightness, an unusually large extraction radius was required for point source removal, which would have significantly impacted the data. Since these observations constituted only a small fraction of the total exposure time (in Ms), we opted to remove them entirely.

Chandra has four science instruments on board. There are two focal plane science instruments, the High Resolution Camera (HRC) and Advanced CCD Imaging Spectrometer (ACIS). There are also two gratings spectrographs that provide high spectral resolution at both low and high energies (LETG and HETG). ACIS uses CCDs with 0.5" imaging resolution and a moderate spectral resolution of 280 eV at 5.9 keV. There are two sets of CCD chips in different layouts: ACIS-I and ACIS-S. ACIS-I (with a 2x2 chip setup) provides the biggest field of view (16.9 x 16.9 arcmin), while ACIS-S is made up of six consecutive chips. All data was cleaned and analyzed using CIAO 4.17.

We created a merged event file with a total exposure time of 1.3 Ms by stacking 80 individual Chandra observations (listed in Table \ref{tab:chandra_data}). To minimize contamination from background flares, we extracted light curves below 2 keV from the outer regions of the CCDs—areas primarily dominated by particle background—and removed time intervals affected by flaring. After flare filtering, we performed astrometric alignment across all observations by matching World Coordinate System (WCS) references using the $\texttt{celldetect}$ tool. Observations with insufficient exposure time for reliable point-source detection were excluded from the final merge. The resulting dataset consists of the OBSIDs listed in Table \ref{tab:chandra_data}, reprojected and co-added with $\texttt{reproject\_obs}$.

For the diffuse emission images, we identified and masked point sources using $\texttt{wavdetect}$ (details provided in Appendix \ref{appendix_sec:stacking}). We generated flux-calibrated images with $\texttt{flux\_obs}$, masked out point sources, and filled the masked regions using $\texttt{dmfilth}$ to achieve a smooth background. Finally, we applied exposure correction by dividing the flux images by their corresponding energy-dependent exposure maps, also produced by $\texttt{flux\_obs}$. The resulting diffuse emission maps are shown in Figure \ref{fig:diffuse-emission}. We note that the effective Chandra exposure and instrumental response have evolved over the mission lifetime, particularly below 2~keV \citep[see, e.g.][]{Montez2024}, such that combining observations taken over two decades can introduce artificial soft features. However, these maps are used solely for visualization and qualitative comparison, and do not enter into the quantitative analysis or scientific interpretation presented here.

\begin{deluxetable*}{c|c|c}
\centering
\tablewidth{0pt}
\tablecaption{Chandra observations used in this work.}
\tablehead{
\colhead{OBSID} & \colhead{Exposure Time (ks)} & \colhead{Start Date} 
}
\startdata
            2953            & 11.59             & 2002-04-19~         \\
    2952            & 11.86             & 2002-03-23~         \\
    2951            & 12.37             & 2002-02-19~         \\
    2954            & 12.45             & 2002-05-07~         \\
    5954            & 17.82             & 2005-08-01~         \\
    13016           & 17.83             & 2011-03-29~         \\
    13017           & 17.83             & 2011-03-31~         \\
    14941           & 19.82             & 2013-04-06~         \\
    14942           & 19.83             & 2013-04-14~         \\
    23295           & 23.37             & 2020-07-03~         \\
    3549            & 24.79             & 2003-06-19~         \\
    22707           & 25.69             & 2020-07-02~         \\
    9170            & 26.80             & 2008-05-06~         \\
    9172            & 27.44             & 2008-05-11~         \\
    9169            & 27.60             & 2008-05-05~         \\
    9171            & 27.69             & 2008-05-10~         \\
    9173            & 27.77             & 2008-07-26~         \\
    9174            & 28.81             & 2008-07-25~         \\
    6363            & 29.76             & 2006-07-17~         \\
    2943            & 37.68             & 2002-05-22~         \\
    3663            & 37.96             & 2002-05-24~         \\
    5951            & 44.59             & 2005-07-27~         \\
    5952            & 45.33             & 2005-07-29~         \\
    5953            & 45.36             & 2005-07-30~         \\
    242             & 45.92             & 1999-09-21~         \\
    22937           & 48.43             & 2020-08-20~         \\
    5950            & 48.53             & 2005-07-24~         \\
    1561            & 49.30             & 2000-10-26~         \\
    4683            & 49.52             & 2004-07-05~         \\
    4684            & 49.53             & 2004-07-06~         \\
    11843           & 78.93             & 2010-05-13~         \\
    3665            & 89.92             & 2002-06-03~         \\
    10556           & 112.55            & 2009-05-18~         \\
    3393            & 158.03            & 2002-05-28~         \\
    3392            & 166.69            & 2002-05-25~         \\\enddata
\tablecomments{List of the \Chandra\ observations used in this work and their respective exposure times and start dates.
}
\label{tab:chandra_data}
\end{deluxetable*}

\section{Data Exploration} \label{sec:data_exploration}

In this section, we examine the diffuse X-ray emission and energy hue maps to establish the physical context for interpreting the components obtained using the pGMCA separation technique. The diffuse emission maps help in identification of energy-dependent morphological structures, while the energy hue maps highlight spectrally distinct structures. Together, these complementary methods characterize both the spatial distribution and spectral properties of the emission components across different energy bands, providing critical benchmarks for validating and interpreting the components derived from pGMCA.

\subsection{Diffuse Emission} \label{sec:diffuse-emission}

\begin{figure*}
    \centering
    \includegraphics[width=0.95\textwidth]{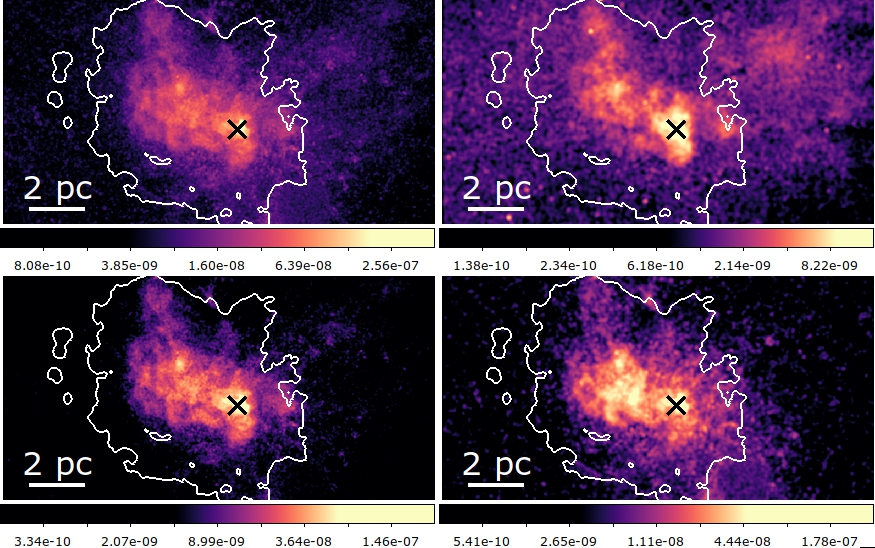}
    \caption{Resulting flux maps (in units of erg/cm$^2$/s) of the stacked 1.3 Ms \Chandra\ ACIS-I observations, listed in Table \ref{tab:chandra_data}. Each image has a 2 pc (50") bar in the bottom left. We have separated the data into the four energy bands of main interest: 2$-$8 keV (full band; top left), 1$-$2.6 keV (Si and S; top right), 2.6$-$5 keV (Ar; bottom left), and 5$-$8 keV (Fe; bottom right). Note that the smoothing scales differ between images. On each image, we plot a contour level at 0.0006 Jy, taken from the radio image \citep{Zhao2016}. The X-ray emission is generally more concentrated than the radio, implying some interaction with surrounding structures.
    }
    \label{fig:diffuse-emission}
\end{figure*}

Following point source removal, we followed the procedures outlined in the CIAO documentation\footnote{\url{https://cxc.harvard.edu/ciao/threads/diffuse-emission/}} to create images of the diffuse X-ray emission. The resulting exposure-corrected images across four energy bands are presented in Figure \ref{fig:diffuse-emission}, each revealing different aspects of the diffuse X-ray emission in the GC region. The top left panel shows the broadest energy band image, with photons between 2$-$8 keV. Here, both the remnant and plasma bound to \SgrA\ are visible, but not distinct. In the 1$-$2.6 keV band (top right panel), we detect faint emission from the wind-fed plasma environment around \SgrA, along with broader field-filling GCXE/GDXE emission and a cloud of soft extended emission to the top right.
The complete absence of the supernova remnant's morphology in this band strongly suggests that the SNR plasma is substantially hotter than the material surrounding \SgrA. Between 2.6$-$5 keV, we start to see a structure emerge to the left of \SgrA. This band contains the majority of the photons in our \Chandra\ observations. Transitioning to higher energies (bottom right panel), the supernova remnant structure becomes more visible, while the plasma around \SgrA\ lacks clear morphology. This aligns with previous studies that characterized the SNR morphology using 6$-$7 keV photons, as the remnant has strong \ironK\ emission \citep{Maeda2002}. However, even in this band, the remnant's morphology remains somewhat ambiguous -- an issue we hope to solve with pGMCA.

\subsection{Energy Hue Maps} \label{sec:energuy_hue_maps}

Next, we developed energy hue maps to help visualize spectral information present in the X-ray data cube. We used \texttt{dmnautilus} to generate a map file identifying regions containing a sufficient number of photon counts, followed by \texttt{statmap} to calculate the per-pixel median energy values. Subsequently, we applied \texttt{aconvolve} to smooth the resulting energy map. The resulting median energy maps were then combined with diffuse X-ray intensity images (not exposure-corrected) to create a composite visualization where hue corresponds to photon energy, while lightness and saturation reflect the raw X-ray brightness in each pixel. Since median energy calculations depend on photon properties rather than detection efficiency, we used raw (non-exposure-corrected) intensities.

\begin{figure}
    \centering
    \includegraphics[width=0.48\textwidth]{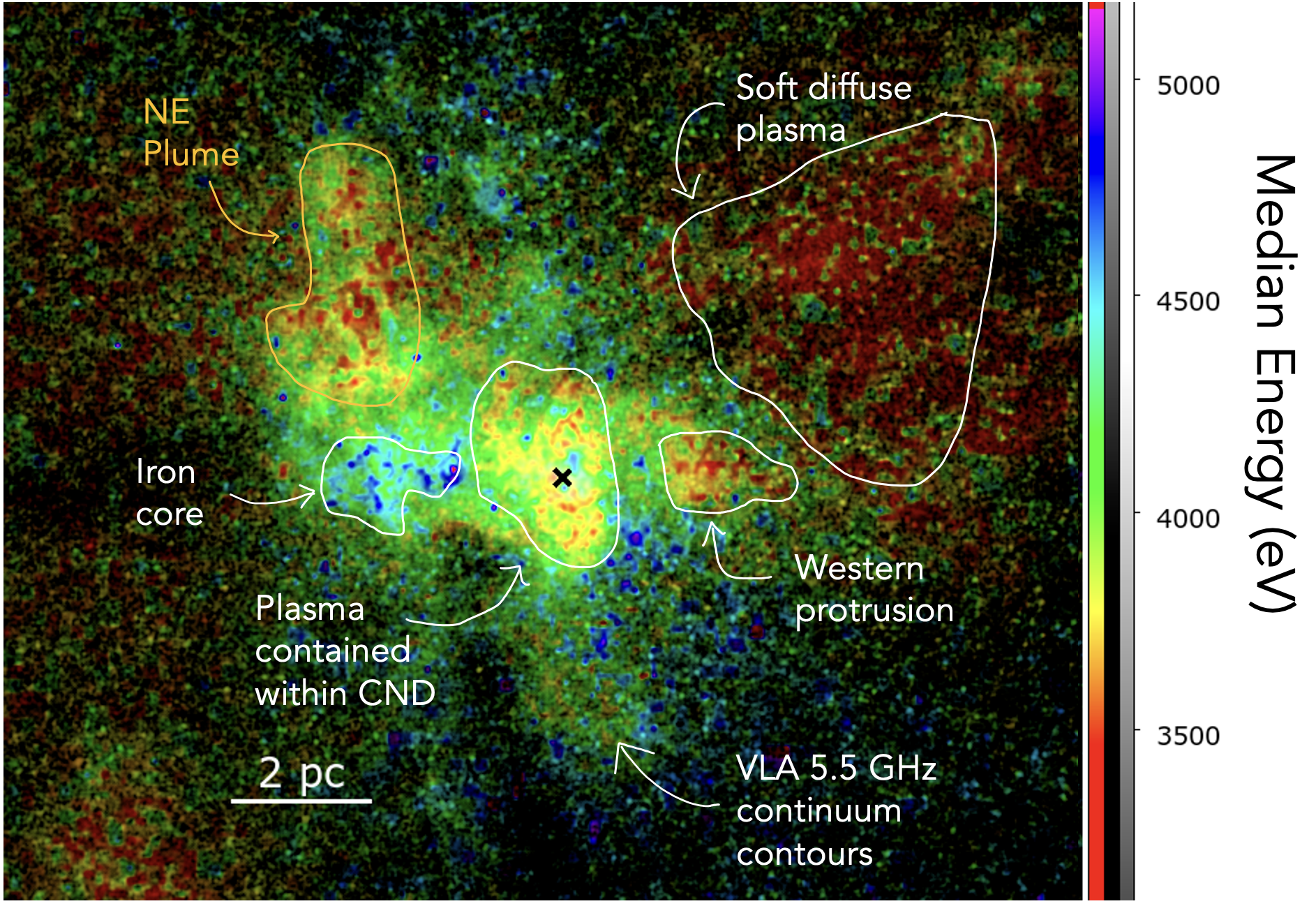}
    \caption{Energy hue map of the region spanning 2$-$8 keV. The color of each pixel is dictated by its median energy (shown in the colorbar), while the lightness and saturation (described in the grayscale bar) are determined by the X-ray intensity in that pixel. Brighter and more saturated regions therefore have a higher X-ray intensity. We identify spectrally differentiated features of interest in this work. Note that unlike the diffuse emission images, these are not exposure corrected; see text for details.
    }
    \label{fig:ehm}
\end{figure}

\begin{figure}
    \centering
    \includegraphics[width=0.48\textwidth]{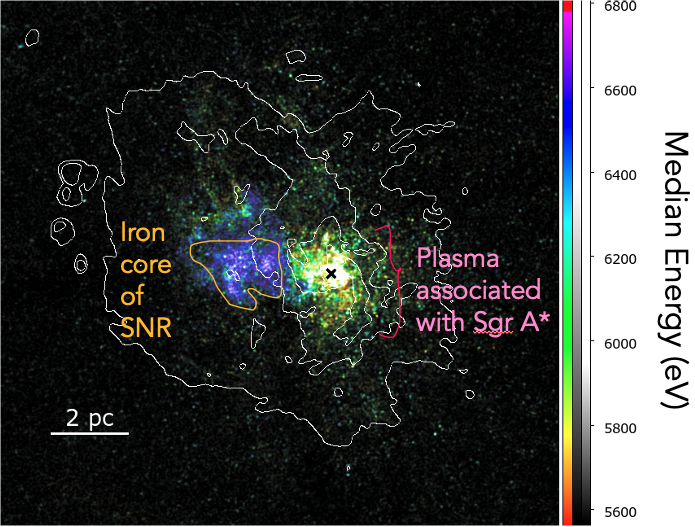}
    \caption{Energy hue maps generated only with photons between 5$-$8 keV. Here, the distinction between the SNR iron core and the plasma surrounding \SgrA\ is highlighted. The plasma associated with \SgrA\ seems to correspond to median energies $\sim$6-6.2 keV while the SNR iron core has median pixel energies closer to 6.7 keV. 
    }
    \label{fig:ehm_5to8}
\end{figure}

The resulting energy hue maps generated using 2$-$8 keV photon events is shown in Figure \ref{fig:ehm}. The left panel highlights several prominent features that stand out in this visualization of the data. There are two areas with redder median pixels that are still X-ray bright. The northeast (NE) plume is brighter than the Western protrusion. Additionally, we label the iron core, which is visually more distinct in Figure \ref{fig:ehm_5to8}. This region has the highest median pixel energies in the whole bright X-ray emission structure. A large, soft diffuse structure extends across the western portion of the image, possibly linked to past activity from \SgrA\ \citep[see, e.g.][]{Maeda2002}. The diffuse plasma near \SgrA\ exhibits median energies softer than those of the iron core but slightly harder than redder X-ray features identified elsewhere in the field. Overlaid in the right panel are contours from the VLA 5.5 GHz continuum image \citep{Zhao2016}, which traces the approximate boundary of the SNRs radio shell. The iron core aligns approximately with the structure's center, and notably, the radio emission is significantly more extended than the X-ray-bright region.

To further investigate the hard X-ray components, we produced an additional energy hue map restricted to 5--8 keV photons (Figure \ref{fig:ehm_5to8}). In this higher-energy band, the iron core of the SNR stands out starkly against the surrounding emission, confirming its significantly harder spectral nature compared to the softer plasma near \SgrA. The spatial-spectral segregation strongly suggests that the SNR emission may be dominated by higher-energy processes, potentially including non-thermal emission or highly ionized plasma, while the \SgrA\ environment appears dominated by cooler thermal emission or distinct physical mechanisms.

\newpage

\section{pGMCA Results} \label{sec:gmca_results}

To separate different plasma components, we used a blind source separation method based on the Poissonian General Morphological Components Analysis \citep[pGMCA, see][]{Bobin2020}, which is based on the original GMCA \citep{bobin2015}. In this section, we briefly describe the algorithm and past applications, followed by our results in applying pGMCA to the point source-removed stacked ACIS-I data cube.

\subsection{Methods} \label{sec:gmca_methods}

GMCA is a blind signal separation technique that works best in separating spectrally distinct components, even if they overlap in our line-of-sight. The original algorithm was designed to extract a CMB image from \textit{Planck} data \citep{Bobin_2016}, and relies on the sparsity of the wavelet transforms of all the $(x,y)$ slices of the $(x,y,E)$ data cube, in order to fully exploit both the three-dimensional nature of the data and the morphological diversity of the components to retrieve. Without any prior physical or instrumental information other than the number $n$ of components to retrieve, the algorithm is able to disentangle $n$ images, each one being associated with a mean spectrum. This disentanglement is made possible by the use of wavelet transforms on the $(x,y)$ slices of the data set, because wavelets carry information on morphological features at different scales that allow the algorithm to separate morphologically distinct components.

This method was first introduced for X-ray observations by \cite{picquenot2019}. An updated version, the pGMCA, was developed to take into account Poisson statistics in \cite{Bobin2020}. The pGMCA algorithm works in a similar way as GMCA, but uses a Poisson likelihood term to ensure data fidelity of the extracted sources instead of assuming linear mixing. However, most matrix factorization techniques
require the data fidelity term to be smooth \citep[][]{xu2015globallyconvergentalgorithmnonconvex}, which is
not the case for the Poisson likelihood: for that reason, the pGMCA uses a smooth approximation of the Poisson likelihood
based on Nesterov’s smoothing technique \citep{Nesterov05}. As a consequence, the components extracted by the pGMCA algorithm are automatically smoothed/denoised. Both algorithms proved their ability to extract detailed and unpolluted maps of faint components from Chandra observations of extended sources \citep[see for example,][]{Picquenot_2021,Olivares_2025}. However, the lack of prior physical information makes the spectral reconstruction challenging, and the algorithm is usually more efficient at disentangling components in narrow energy bands. pGMCA's capabilities make it a great candidate to study the GC and \SgrA. 

In this context, ``narrow energy bands'' refer to intervals over which the relative spectral weights of different plasma components vary slowly, allowing the emission to be approximated as a linear mixture of a small number of morphologically distinct components. Under these conditions, the pGMCA assumption of fixed spectral signatures within a band is well satisfied, and component separation is driven primarily by morphology rather than by spectral curvature. 

In practice, applying pGMCA to Galactic Center data requires a compromise between algorithmic optimality and photon statistics. Although the method is formally optimized for narrower bands, the strong absorption and large surface-brightness contrasts in this region necessitate wider bands to achieve adequate signal-to-noise. We therefore apply pGMCA over the 2--5~keV and 5--8~keV ranges, which preserve sufficient photon counts while limiting spectral complexity to a small number of dominant plasma components. Within these bands, pGMCA robustly isolates morphologically and physically distinct structures that can be interpreted and modeled in subsequent spectral analysis.

\subsection{Outputs} \label{sec:gmca_outputs}

We applied pGMCA to the stacked, ACIS-I observations with point sources removed. The Chandra ACIS detector has a pixel scale of 0.492 arcsec and a full width at half maximum (FWHM) energy resolution of approximately 110 eV at 6.7 keV. To account for the physical limitations of the instrument, we selected binning parameters that balance spectral and spatial resolution. Specifically, we created a data cube binned spectrally in $\sim$10 eV bins and spatially by a factor of 8 (spatial resolution of $\sim$4 arcsec). This binning scheme preserved key spectral features while minimizing noise. As noted in the previous section, pGMCA performs optimally when applied to narrow energy intervals and is less effective across broad energy bands. We found that running the algorithm separately on the 2–5 keV and 5–8 keV bands yielded robust component separation, whereas narrower energy ranges were limited by insufficient signal-to-noise. 


\begin{figure*}
    \centering
    \includegraphics[width=0.98\textwidth]{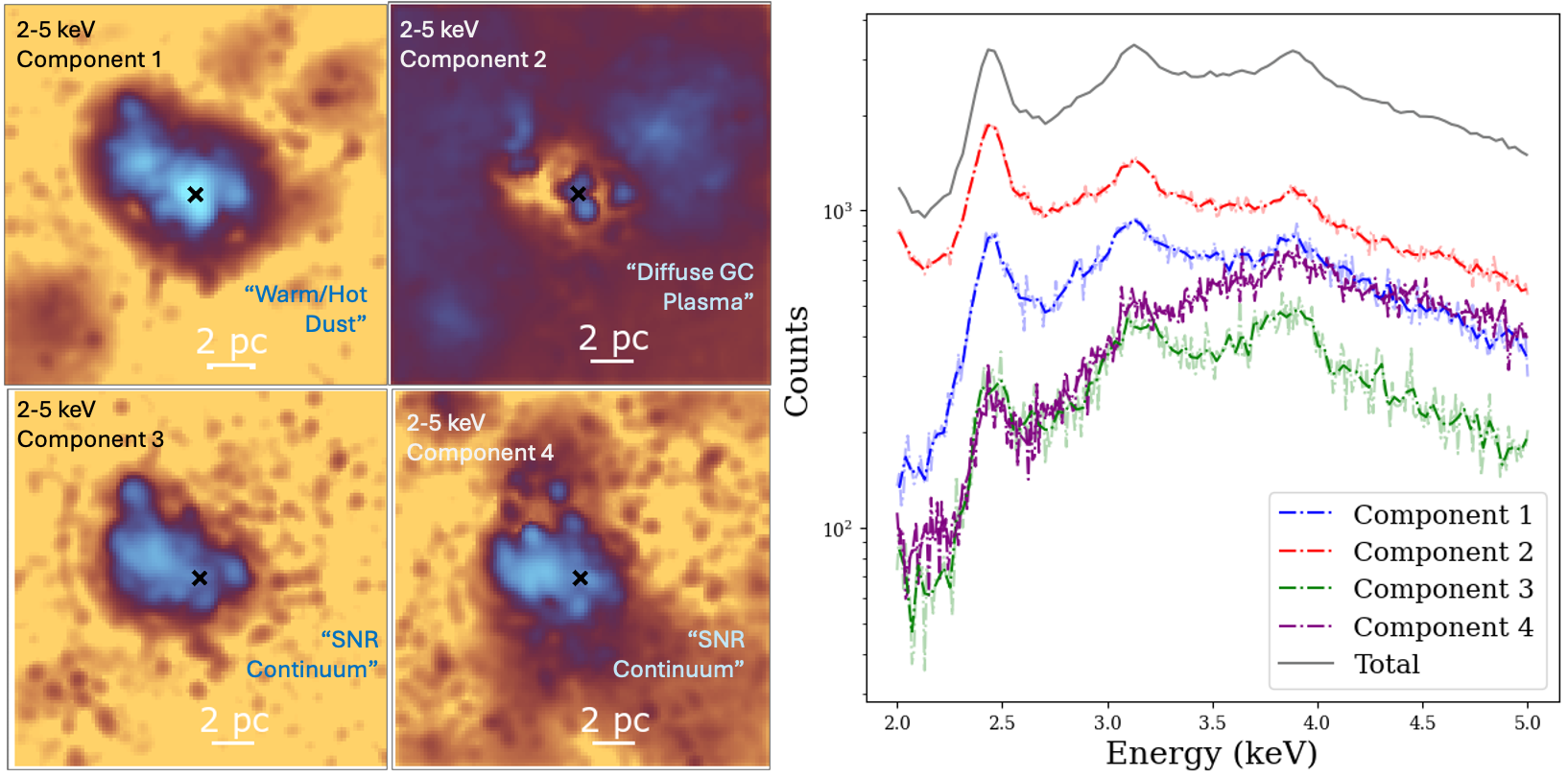}
    \includegraphics[width=0.98\textwidth]{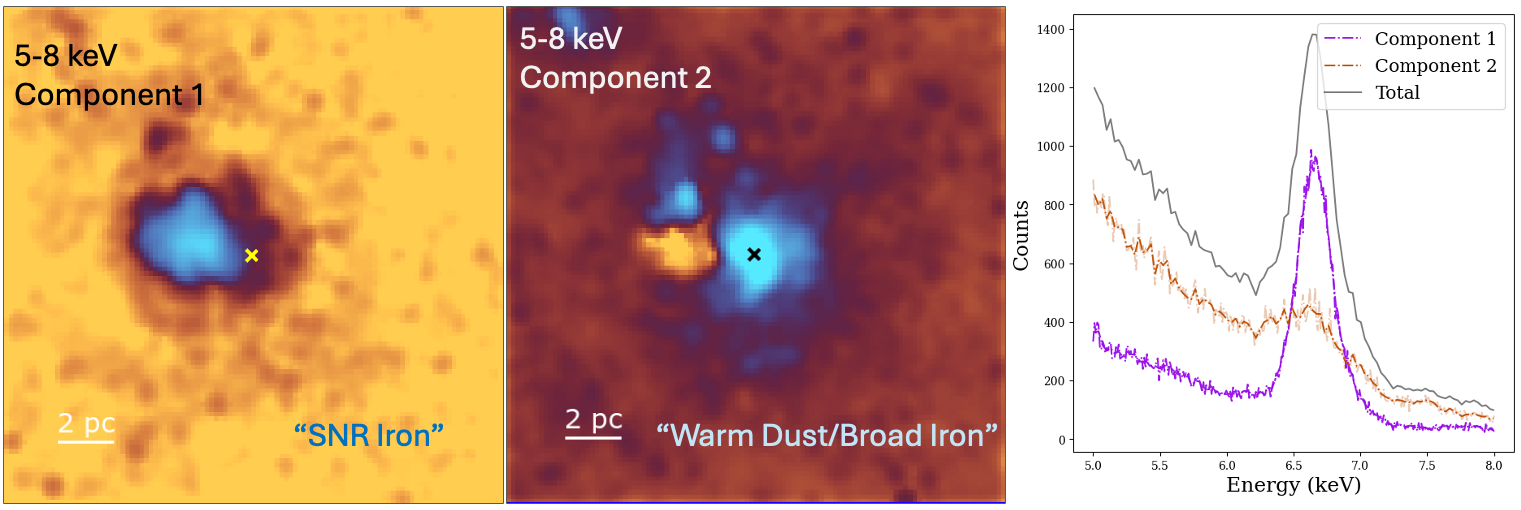}
    \caption{GMCA outputs and corresponding spectra. The output images have been divided by exposure maps created with \texttt{flux\_obs}. All images share the same log-spaced colorbar, with yellow indicating the lowest emission levels and blue corresponding to regions of high emission. The top row shows the two main components seen between 5$-$8 keV, while the top two rows display the four components identified by the algorithm in the 2$-$5 keV energy range. Each component is labeled with notation we refer to throughout this work. }
    \label{fig:gmca_results}
\end{figure*}

Figure \ref{fig:gmca_results} shows our results from applying pGMCA to photons between 5$-$8 keV (top row) and 2$-$5 keV (middle and bottom row). Each image has a scale bar denoting the extent of 2 pc (50 arcseconds at 8.1 kpc) and an X at the position of \SgrA. Each highlighted spectrum (bolded color) is shown next to its morphological counterpart. The total spectrum in the respective energy band is also plotted in dark gray. We omit the noise components for brevity. Note that we have exposure corrected these images; we took the output pGMCA components and divided them by exposure maps binned to match the input pGMCA data cube at the corresponding energies. As noted above, combining exposure maps from different \Chandra\ cycles can introduce artificial features, primarily below 2~keV; however, because the pGMCA decomposition is performed on the 2--5~keV cube, this effect does not significantly impact the results shown here and primarily serves to improve visual clarity. We briefly note features of the images and spectra below, then compare them to multiwavelength images to interpret their physical significance in the following section.

\textbf{5$-$8 keV components:} Application of pGMCA to the 5$-$8 keV band cleanly isolates the 6.7 keV Fe-K$\alpha$ line emission (Component 1), confirming its primary origin in Sgr A East rather than astrophysical processes associated with \SgrA. The center and extent of this feature is coincident with the higher energy emission seen in the 5$-$8 keV energy hue map (Figure \ref{fig:ehm_5to8}) and the 5$-$8 keV diffuse emission (Figure \ref{fig:diffuse-emission}). The next component (iron Component 2) exhibits negligible emission in the SNR core, and the Fe-K$\alpha$ line profile displays a slight energy offset ($\sim$30 eV), possibly indicative of a lower-ionization plasma. The shape of the plasma surrounding \SgrA\ retrieved by pGMCA is morphologically very similar to the 1$-$2.6 keV plasma surrounding \SgrA\ seen in Figure \ref{fig:diffuse-emission}.


\textbf{2$-$5 keV components:}  Component 1 shows emission co-spatial with the two redder blobs of softer gas from the hue map (Figure \ref{fig:ehm}). Stronger emission appears to be from the plasma environment of \SgrA.  Component 2 is the brightest component and displays a prominent sulfur line. Its morphology is anti-correlated with that of the supernova remnant and appears co-spatial with the molecular filament. This component also has emission co-spatial with the NSC, but it is much less extended than in Component 1, and seems to be contained within the  CND. Both Components 1 and 2 display vertical blobs extending perpendicular to the Galactic plane, previously interpreted as potential signatures of past ejections from \SgrA\ \citep{Maeda2002}. Component 3 exhibits a bright core located near the center of the radio shell, suggesting an association with the supernova remnant. Its spectrum closely resembles that of Component 4 but is distinguished by the presence of a calcium emission line. Component 4 appears more compact than Component 3, but is similarly co-spatial with both SNR core and NSC. Comparing to Component 3, its spectral features are weaker. We ran the pGMCA algorithm with $n = 4$ components between 2-5 keV and with $n = 2$ between 5-8 keV. When running it with $n = 3$ components in the 2-5 keV range, the algorithm combines Components 3 \& 4, whereas if we run the algorithm with $n = 5$ components, it will produce a noise component.

We fit the spectrum of each pGMCA component to measure the line centroids and full widths at half maximum (FWHM), allowing us to compare the components and draw conclusions about the physical processes at play. We calculated the line centroids by fitting Voigt profiles to the continuum-subtracted spectra produced by pGMCA, using tools from \texttt{specutils} and \texttt{astropy.modeling}. The fits were optimized with the Levenberg–Marquardt algorithm with a least-squares minimization. Reported uncertainties correspond to $3\sigma$ errors, derived from the diagonal elements of the covariance matrix returned by each fit.

Table \ref{tab:line-centroids} gives the line centroids, corresponding FWHMs, and elemental transitions that could plausibly give rise to these lines. We note that fitting Component 4's spectrum was particularly challenging; the algorithm had difficulty identifying peaks due to the lower strength of the 3.1 and 3.9 keV features. As a result, the derived line centroids and widths for this component should be interpreted with caution.

\begin{deluxetable*}{lllll}
\centering
\tablewidth{0pt}
\tablecaption{Line Centroids and full width half maxes from fitting pGMCA outputs and potential associated ions.}
\tablehead{
\colhead{Energy Band} & \colhead{pGMCA Component} & \colhead{Line FWHM (keV) } & \colhead{Line Centroid (keV)} & \colhead{Possible Ions} 
}
\startdata
2--5 keV & 1 & 0.196 & 2.460 $\pm$ 0.008 & S XV$^a$\\
        \hline
2--5 keV & 1 & 0.234 & 3.115 $\pm$ 0.014 & Ar XVII$^b$ or S XVI$^c$ \\
        \hline
2--5 keV & 1 & 0.084 & 3.920 $\pm$ 0.032 & Ca XIX$^d$\\
        \hline
2--5 keV & 2 & 0.219 & 2.495 $\pm$ 0.007 & S XV$^a$ or Si XIV$^e$ \\
        \hline
2--5 keV & 2 & 0.253 & 3.114 $\pm$ 0.009 & Ar XVII$^b$ or S XVI$^c$\\
        \hline
2--5 keV & 2 & 0.209 & 3.891 $\pm$ 0.014 & Ca XIX$^d$\\
        \hline
2--5 keV & 3 & 0.044 & 2.438 $\pm$ 0.016 & S XV$^a$ \\
        \hline
2--5 keV & 3 & 0.064 & 3.126 $\pm$ 0.033 & Ar XVII$^b$ or S XVI$^c$ \\
        \hline
2--5 keV & 3 & 0.077 & 3.860 $\pm$ 0.036 & Ca XIX$^d$\\
        \hline
2--5 keV & 4 & 0.194 & 2.369 $\pm$ 0.032* & Si XIV$^e$, Si XIII$^f$, or S XV$^a$ \\
        \hline
2--5 keV & 4 & 0.037 & 3.148 $\pm$ 0.030 & Ar XVII$^b$ or S XVI$^c$\\
        \hline
2--5 keV & 4 & 0.051 & 3.905 $\pm$ 0.043 & Ca XIX$^d$\\
\hline
5--8 keV & 1 & 0.186 & 6.630 $\pm$ 0.028 & Fe XXV$^g$ or Fe XXIV$^h$ \\
\hline
5--8 keV & 2 & 0.460 & 6.663 $\pm$ 0.004 & Fe XXV$^g$ or Fe XXIV$^h$ \\
\hline
\multicolumn{5}{l}{a: S XV (He-like): R: 2.461 keV; I: 2.447, 2.449 keV; F: 2.430 keV; kT$_{p}$ = 1.289 keV} \\[4pt]
\multicolumn{5}{l}{b: Ar XVII (He-like): R: 3.140 keV; I: 3.124 keV; F: 3.104 keV; kT$_{p}$ = 1.821 keV} \\[4pt]
\multicolumn{5}{l}{c: S XVI (H-like): R: 3.107, 3.106 keV; kT$_{p}$ = 2.293 keV} \\[4pt]
\multicolumn{5}{l}{d: Ca XIX (He-like): R: 3.902 keV; F: 3.861 keV; kT$_{p}$ = 2.573 keV} \\[4pt]
\multicolumn{5}{l}{e: Si XIV (H-like): R: 2.377, 2.376 keV; kT$_{p}$ = 1.447 keV} \\[4pt]
\multicolumn{5}{l}{f: Si XIII (He-like): R: 2.374, 2.346 keV; kT$_{p}$ = 0.967 keV} \\[4pt]
\multicolumn{5}{l}{g: Fe XXV (He-like): R: 6.700 keV; I: 6.682, 6.668 keV; F: 6.637 keV; kT$_{p}$ = 5.759 keV} \\[4pt]
\multicolumn{5}{l}{h: Fe XXIV (Li-like): R: 6.682 keV; kT$_{p}$ = 3.849 keV} \\
\enddata
\tablecomments{Line features identified in the 2--5 keV band across pGMCA components. Candidate lines are matched by centroid energy and temperature peak ($kT_p$) to known transitions. Errors on the values are 3$\sigma$ uncertainties. Below, we list the possible ions corresponding to the spectral features in the pGMCA outputs, with the recombination (R), intercombination (I), and forbidden (F) line centroids, along with peak plasma temperatures associated with the resonance lines. }
\label{tab:line-centroids}
\end{deluxetable*}

\subsection{The Sgr A East SNR} \label{sec:snr-structure}

\begin{figure*}
    \centering
    \includegraphics[width=0.98\textwidth]{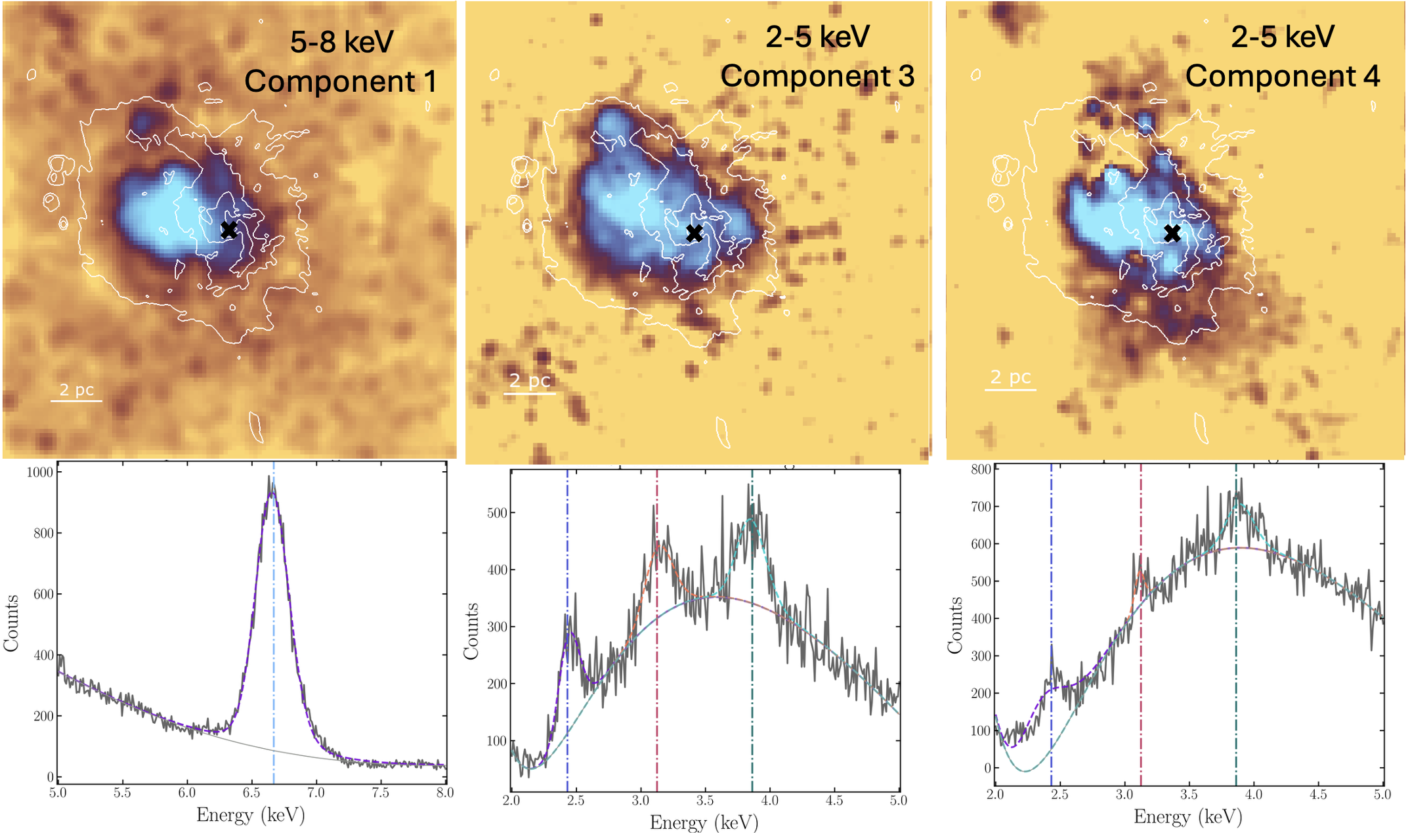}
    \caption{GMCA outputs we believe capture emission from the supernova remnant. Each image has the radio shell \citep{Zhao2016} overlaid in white, an X at the location of \SgrA, and a scale bar showing the extent of 2 parsecs in the bottom left. Below each image, we plot the associated spectrum. The panel on the left is from the 5$-$8 keV component, while the other two are emission components from the 2$-$5 keV data. The line centroids and widths can be found in Table \ref{tab:line-centroids}. }
    \label{fig:gmca_snr}
\end{figure*}

As seen in the previous section, applying pGMCA to the stacked and point source removed \Chandra\ dataset has pulled out several components that appear centered in the middle of the 5.5 GHz radio shell \citep{Zhao2016}. Figure \ref{fig:gmca_snr} shows the components from Figure \ref{fig:gmca_results} that we attribute to the supernova remnant, overlaid with the 5.5 GHz continuum VLA contours in white \citep{Zhao2016}. The images have the same colorbar and limits. 

Examining the pGMCA spectra and morphologies of three key components in Figure \ref{fig:gmca_snr} reveals their distinct physical origins. The 5$-$8 keV component forms a compact, uniformly bright core reminiscent of the isothermal centres seen in mixed-morphology supernova remnants. Its spectrum is dominated by a strong \ironK\ line at 6.663 $\pm$ 0.004 keV with a full width at half maximum of 0.46 keV. By contrast, the 2$-$5 keV component 3 extends farther to the north and west and exhibits a broader, less concentrated core whose peak is offset from the iron centre. Its spectrum shows comparably strong sulfur, argon and calcium lines, consistent with a multi-temperature plasma composed of supernova ejecta, and the spatial offset suggests element-dependent mixing within the remnant. The 2$-$5 keV component 4 isolates the lower-energy continuum underlying the iron emission. The corresponding spectrum exhibits weak spectral features compared to the other pGMCA outputs.

\begin{figure}
    \centering
    \includegraphics[width=0.48\textwidth]{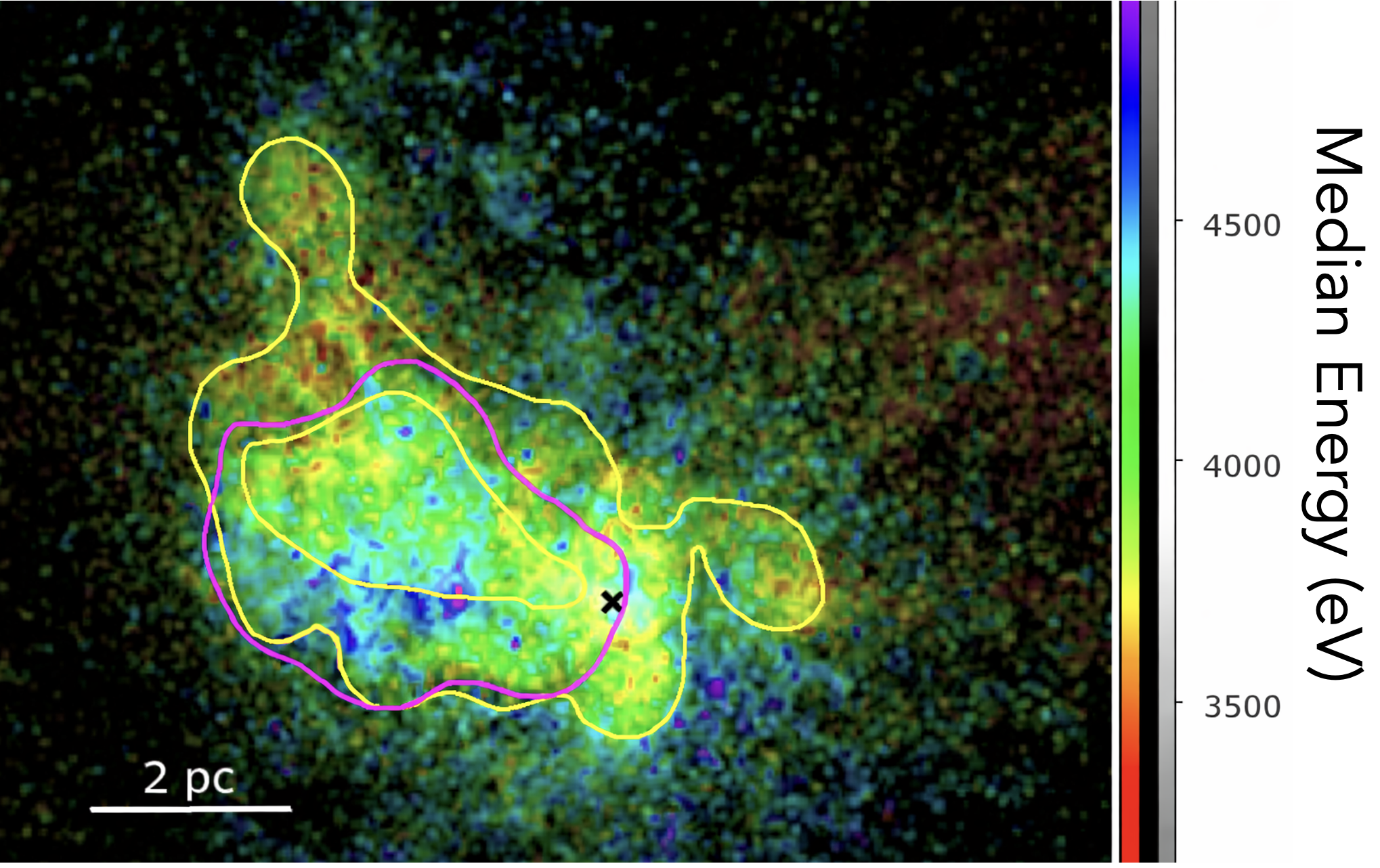}
    \caption{Median energy map of 2$-$8 keV photons (identical to Figure \ref{fig:ehm}). The brightness and saturation in each pixel correspond to the X-ray intensity, while the color corresponds to the median energy (colorbar in units of eV). Overlaid in pink and yellow are the left and middle panels, respectively, from Figure \ref{fig:gmca_snr}. The 2$-$5 keV ejecta component is co-spatial with the two X-ray bright and red features seen in Figure \ref{fig:ehm}. The iron line component appears to be centered on high median energy pixels, the iron core denoted in Figure \ref{fig:ehm}. }
    \label{fig:ehm_snr}
\end{figure}

To see how this compares to the X-ray emission overall, in Figure \ref{fig:ehm_snr}, we plot the pGMCA contours from the left and middle panels in Figure \ref{fig:gmca_snr} in pink and yellow, respectively. The line-dominated 2$-$5 keV component is co-spatial with the NE plume and Western protrusion described earlier (Section \ref{sec:energuy_hue_maps}), implying the supernova remnant has extended up and to the east. The \ironK\ line emission is more concentrated, and looks centered on the high value median energy pixels in the previously denoted iron core.

To contextualize the interaction between Sgr A East and its surrounding environment, we compare our results to multiwavelength datasets capturing emission between 10 K and 50 K. Adjacent to Sgr A East lies a molecular filament, historically divided into distinct components: the 50 $\mathrm{km \, s^{-1}}$ cloud to the east and the 20 $\mathrm{km \, s^{-1}}$ cloud to the south. However, this structure forms a continuous filament connected by magnetic field lines (see panel A in Figure \ref{fig:multiwavelength}) and exhibits a smooth velocity gradient, decreasing from 50 $\mathrm{km \, s^{-1}}$ in the southwest to 20 $\mathrm{km \, s^{-1}}$ in the south \citep{Coil2000}. This filament is traced by SCUBA 450 \micron\ emission (Figure \ref{fig:multiwavelength}, panel A), which reveals cold dust at temperatures of 10–20 K \citep{Pierce-Price2000}, while HI-Gal derived dust maps (PPMAPs; Figure \ref{fig:multiwavelength}, panels B-D) resolve distinct dust structures at intervals between 10 and 50 K \citep{Marsh2017}. Surveys of OH 1720 MHz and Class I methanol (44 GHz) masers detect numerous masers along the filament. Those in the 20 $\mathrm{km \, s^{-1}}$ cloud are associated with ongoing star formation, while masers near 0 and 50 $\mathrm{km \, s^{-1}}$ likely trace shocks or cloud-cloud collisions \citep{Yusef-Zadeh1999, Pihlstrom2006}.  

\begin{figure*}
    \centering
    \includegraphics[width=0.98\textwidth]{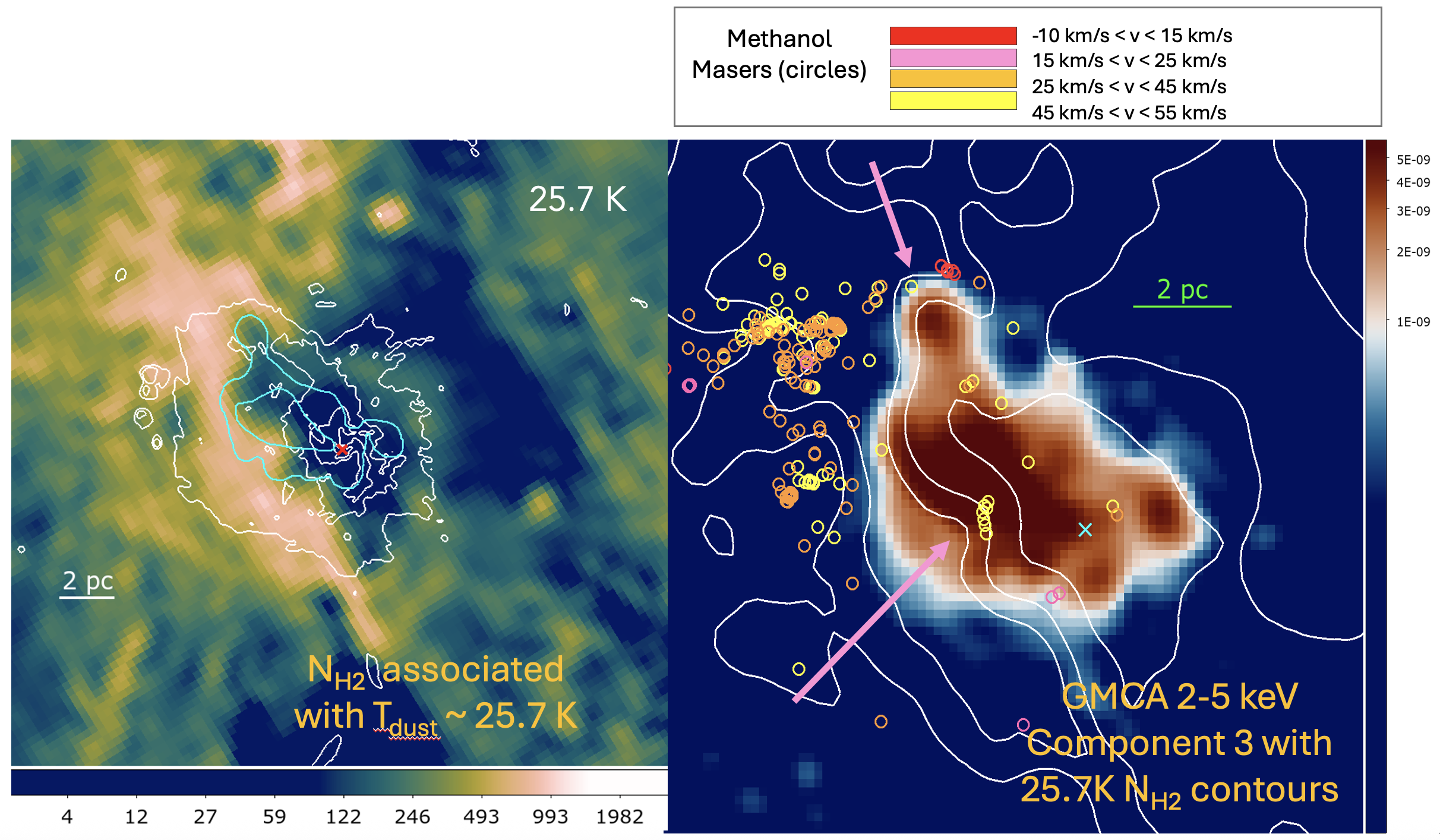}
    \caption{LEFT: $N_{H_2}$ column densities of gas associated with dust at 25.7 K \citep[derived from HI-Gal plane survey dust maps][]{Marsh2017}, with VLA 5.5 GHz contours and 2$-$5 keV pGMCA Component 3 contours in white and cyan, respectively. RIGHT: pGMCA 2$-$5 keV component 3, from Figure \ref{fig:gmca_snr}, with 44 GHz methanol masers \citep{McEwen2016} plotted (maser legend is to the left) and contours from the 25.7 K emission. The pink arrows in the bottom right image highlight regions where it appears the X-ray plasma is pushing up against material associated with 25.7 K dust.  }
    \label{fig:snr_masers_sio}
\end{figure*}

In Figure \ref{fig:snr_masers_sio}, we compare pGMCA 2$-$5 keV Component 3 (right) to $N_{H_2}$ column density maps corresponding to dust with $T_{dust} \sim 25.7 \ \rm K$ (left). The colorbars have been chosen to highlight contrast. Note that gaps in the column density do \textit{not} imply a lack of dust, in fact, the material surrounding \SgrA\ contains dust closer to 30-40 K (see Figure \ref{fig:multiwavelength}, panel D). We also overplot 44 GHz Class I Methanol masers \citep{McEwen2016}; the legend explaining the color scheme is to the left of the image. On the pGMCA image, we have overlaid contours corresponding to the $N_{H_2}$, and highlighted with pink arrows areas of interest. Both arrows are pointing to areas where it is clear that the X-ray emission is being compressed against material, preventing it spreading further. The dust in this molecular material is at 25.7 K, warmer than the dust associated with the 20 $\mathrm{km \, s^{-1}}$ material to the south (see Figure \ref{fig:multiwavelength}). Along with previous maser observations \citep[see, e.g.][]{McEwen2016} which cluster along the X-ray ridge, indicating regions of enhanced density and shock excitation, this implies the remnant's expansion has heated the 50 $\mathrm{km \, s^{-1}}$ cloud and compressed dust. 

The bottom pink arrow marks the sharp interface between the pGMCA component and the 25.7 K dust, delineating the reflected-shock front. At this boundary, ejecta heated by the reflected shock has been compressed and thermalized, creating the nearly isothermal core. Masers at both 0 $\mathrm{km \, s^{-1}}$ and 50 $\mathrm{km \, s^{-1}}$ cluster along this edge, confirming that the reflected shock propagates simultaneously toward and away from us along our line of sight. 

\begin{figure*}
    \centering
    \includegraphics[width=0.98\textwidth]{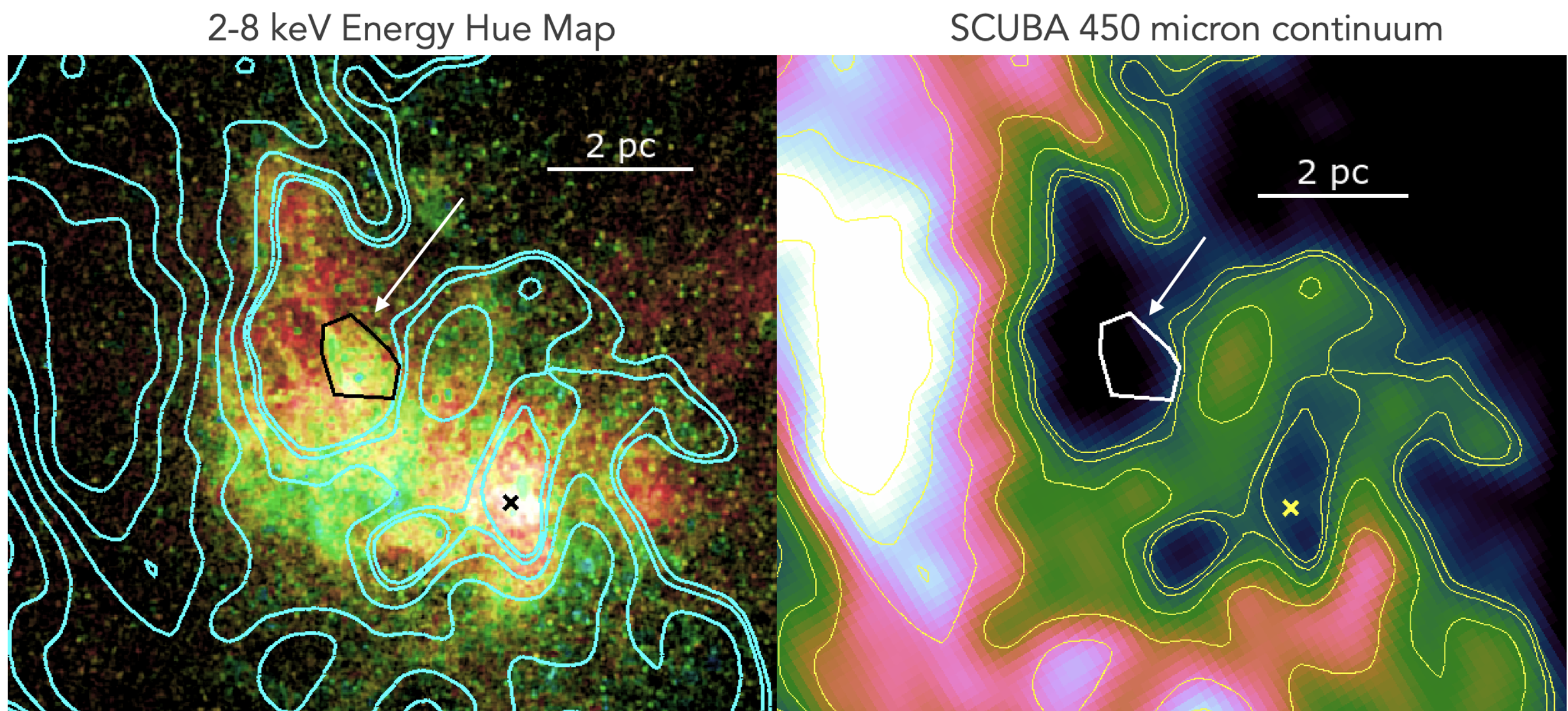}
    \caption{LEFT:  2--8 keV energy hue map, with the contours from the right panel shown in cyan.
RIGHT: SCUBA 450 \micron\ emission (colormap) with overlaid contours. Arrows mark gaps in the molecular material, and the red regions in the energy map correspond precisely to these holes. This alignment indicates that the SNR lies behind the molecular gas where the median pixel energy is higher, since the softer X-ray emission has been absorbed by the foreground material.
}
    \label{fig:snr_molecular}
\end{figure*}

In Figure \ref{fig:snr_molecular}, we show the energy hue map on the left and SCUBA 450 \micron\ continuum map on the right. The cyan contours trace the dense molecular gas. In the left panel, the color delineates the median energy of the pixel. Overlaying the SCUBA contours reveals that red, X-ray bright regions align precisely with cavities devoid of cold (10$-$20 K) and dense molecular gas. The spatial correlation between cold dust and hard X-ray regions also implies that greener areas in the hue map trace molecular material lying in front of the remnant. This foreground gas absorbs soft X-rays, thereby raising the median photon energy. 


\subsection{Plasma associated with the NSC and/or Warm Dust} \label{sec:nsc_dust}

\begin{figure*}
    \centering
    \includegraphics[width=0.98\textwidth]{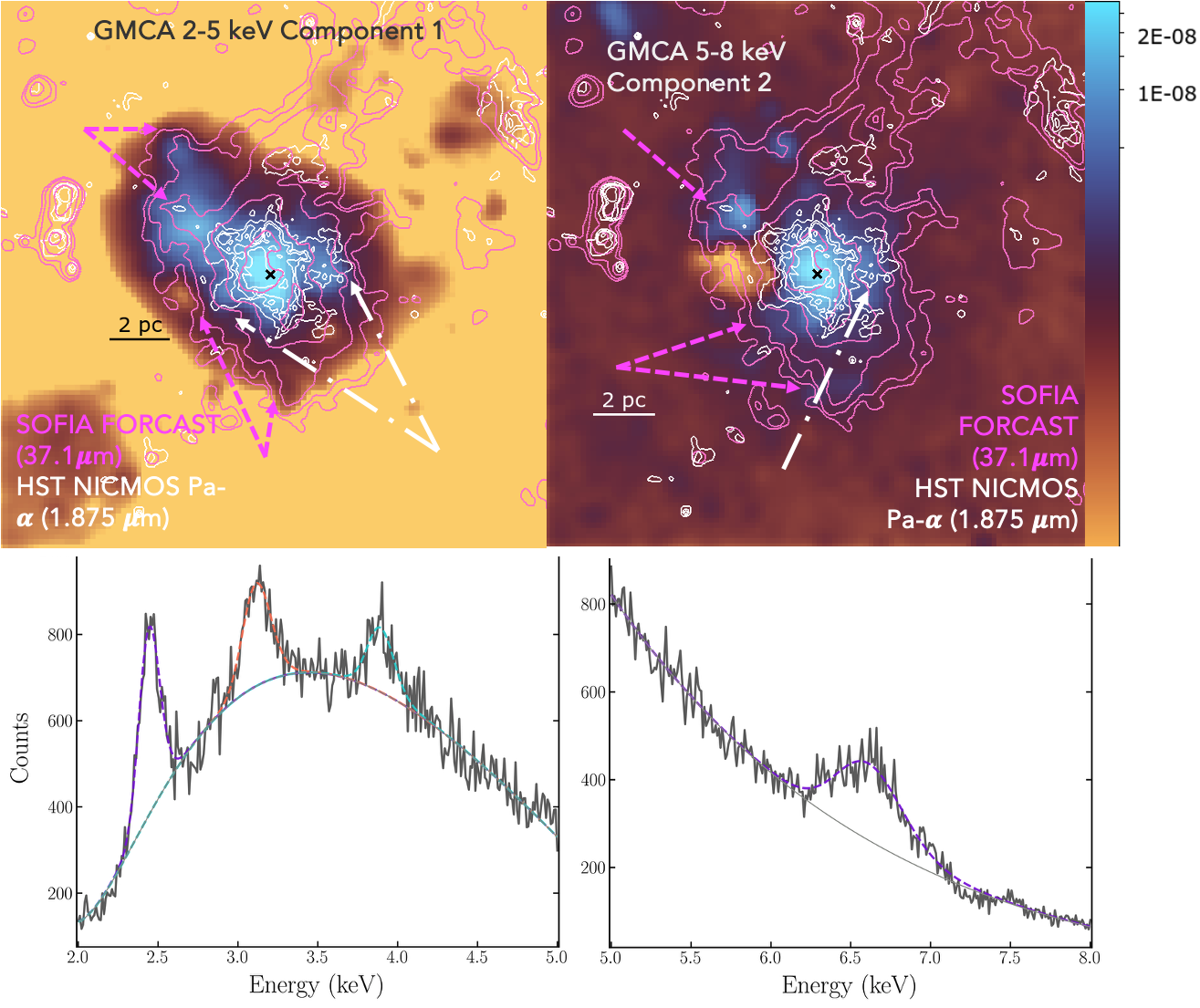}
    \caption{GMCA outputs that appear co-spatial with warm and hot gas emission. Each image has cnotours from 37.1\micron\ $\sim$ SOFIA FORCAST dust emission (T$_{dust} \sim 100 K$) and 1.875 \micron\ Pa-$\alpha$ HST NICMOS $\sim$ (T$_{gas} \sim 100 K$) emission in pink and white, respectively. They also have an X at the location of \SgrA\ and a scale bar showing the extent of 2 parsecs in the bottom left. The panel on the left is emission between 2$-$5 keV while the panel on the right is emission between 5$-$8 keV. The pink and white arrows denote regions where the contours align particularly well with the pGMCA extracted component. The corresponding spectra are plotted below each image, with continuum and line fits that were used in the calculation of line centroids listed in Table \ref{tab:line-centroids}. The component on the left traces hot plasma pockets that coincide with both the NSC and dust within the Sgr A East shell \citep{Lau2015}. This implies that the dust (denoted by the right arrow) has endured the passage of the reflected shock and has been shock heated.}
    \label{fig:gmca_dust}
\end{figure*}

We identify two components that are spatially coincident with warm dust and hot gas features (Figure \ref{fig:multiwavelength}; panels F and G). Figure \ref{fig:gmca_dust} shows images and spectra of pGMCA 2$-$5 keV Component 1 and 5$-$8 keV Component 2, with HST NICMOS Pa-$\alpha$ 1.875 \micron\ (hot gas) and SOFIA Forcast 37.1 \micron\ (warm dust; T$_{dust} \sim 100 K$) contours overlaid in white and pink, respectively. The arrows indicate locations where the contours of the corresponding color align with the pGMCA-extracted component, highlighting spatial coincidence. In the left panel, the X-ray emission is co-spatial with the warm dust in the NE plume and the hot gas protruding to the east and west of \SgrA. The 5$-$8 keV component is co-spatial with the warm dust in the NE plume and hot gas in the western direction, but not the extension to the east. 

Looking at the spectra, the 2$-$5 keV component has three features at 2.46 $\pm$ 0.008 keV (S~\textsc{xv}), 3.115 $\pm$ 0.014 keV (could be Ar~\textsc{xvii} at 3.124 or 3.104 keV, or S~\textsc{xvi} at 3.106 or 3.107 keV), and $3.920 \pm 0.032$ keV (Ca~\textsc{xix}). The FWHMs for the first two lines are $\sim$ 0.2 keV, while the Calcium FWHM is closer to 0.01 keV. The 3.115 keV line centroid is in between the two strongest possible lines, the Ar~\textsc{xvii} 5-1 transition at 3.124 keV and the Ar~\textsc{svii} 2-1 transition at 3.104 keV, both peaking in plasmas at $kT_e \sim1.7$ keV. It could also be lines associated with S~\textsc{xvi} closer to 3.107 keV, which would correspond to $kT_e \sim 2.3$ keV. As the S~\textsc{xv} feature is strongest, corresponding to temperatures between $kT_e =1.22 - 1.29$ keV, the plasma is likely closer to the lower range $\sim 1.2-1.3$ keV. Meanwhile, the 5$-$8 keV component has a broad iron line centered at 6.63 $\pm$ 0.03 keV and a FWHM of 0.186 keV. The iron line here is weaker and slightly redshifted (by $\sim$ 30 eV) compared to the iron core (5$-$8 keV Component 1), indicating lower abundance or temperature. 

The warm dust emission observed by SOFIA Forcast, specifically the region highlighted with a white arrow in Figure \ref{fig:gmca_dust}, was analyzed in previous work by \citealt{Lau2015}. In that work, the authors found that this region of dust emission exhibited an average temperature of 100 K, significantly higher than the 75 K expected for typical 0.1 \micron\ ISM grains heated by the NSC, implying a population of smaller grains (0.01 \micron). They postulated that the dust was located \textit{within} Sgr A East due to a spatial distribution anti-correlated with hard X-ray (2-8 keV) emission, suggesting a cooler, denser ejecta region. The absence of cold dust and ionized gas emission further excluded a scenario in which the dust was in a foreground cloud. SED modeling incorporating SOFIA and Spitzer data required a mix of very small grains (VSGs, $\sim$0.001 \micron) and larger grains (LGs, $\sim$0.04 \micron), with a VSG-to-LG mass ratio (15–90\%) exceeding typical ISM values (13\%). Compositional constraints ruled out PAHs, favoring amorphous carbon, and derived a dust mass of about 0.02 M$_\odot$.

The 2$-$5 keV component in particular (Figure \ref{fig:gmca_dust}; left) aligns precisely with the ionized gas protruding from \SgrA/NSC, identifying these blobs as hot pockets of plasma. This ionized material likely originates in the NSC, but it is unclear why it is extending perpendicular to the Galactic plane, instead of following the plume in the northwest.

As mentioned, both of these pGMCA components are co-spatial with the dust component characterized by \citealt{Lau2015} (white arrow in Figure \ref{fig:gmca_dust}). However, no corresponding X-ray emission is detected from the H II regions to the east of the remnant, or the northwestern protrusion visible in the dust contours and the radio contours (e.g. Figure \ref{fig:multiwavelength}).  This difference strongly suggests two distinct heating mechanisms: the dust that overlaps X-ray emission has been shock-heated by the supernova blast wave, whereas the dust embedded in H II and in the NW protruding structure is warmed predominantly by stellar UV photons. Additionally, the NE plume and Western protrusion (labeled in Figure \ref{fig:ehm}) were both seen in pGMCA Component 3 (Figure \ref{fig:gmca_snr}; middle), which we attribute to ejecta/swept up material from the remnant. Together with the fact that these H II regions lie outside the current reflected-shock radius (Section \ref{sec:snr-structure}), our findings bolster \citet{Lau2015}'s conclusion that this dust resides inside the remnant shell and has survived the passage of the reflected shock.

\subsection{Galactic Center X-ray Emission} \label{sec:diffuse}

\begin{figure*}
    \centering
    \includegraphics[width=0.98\textwidth]{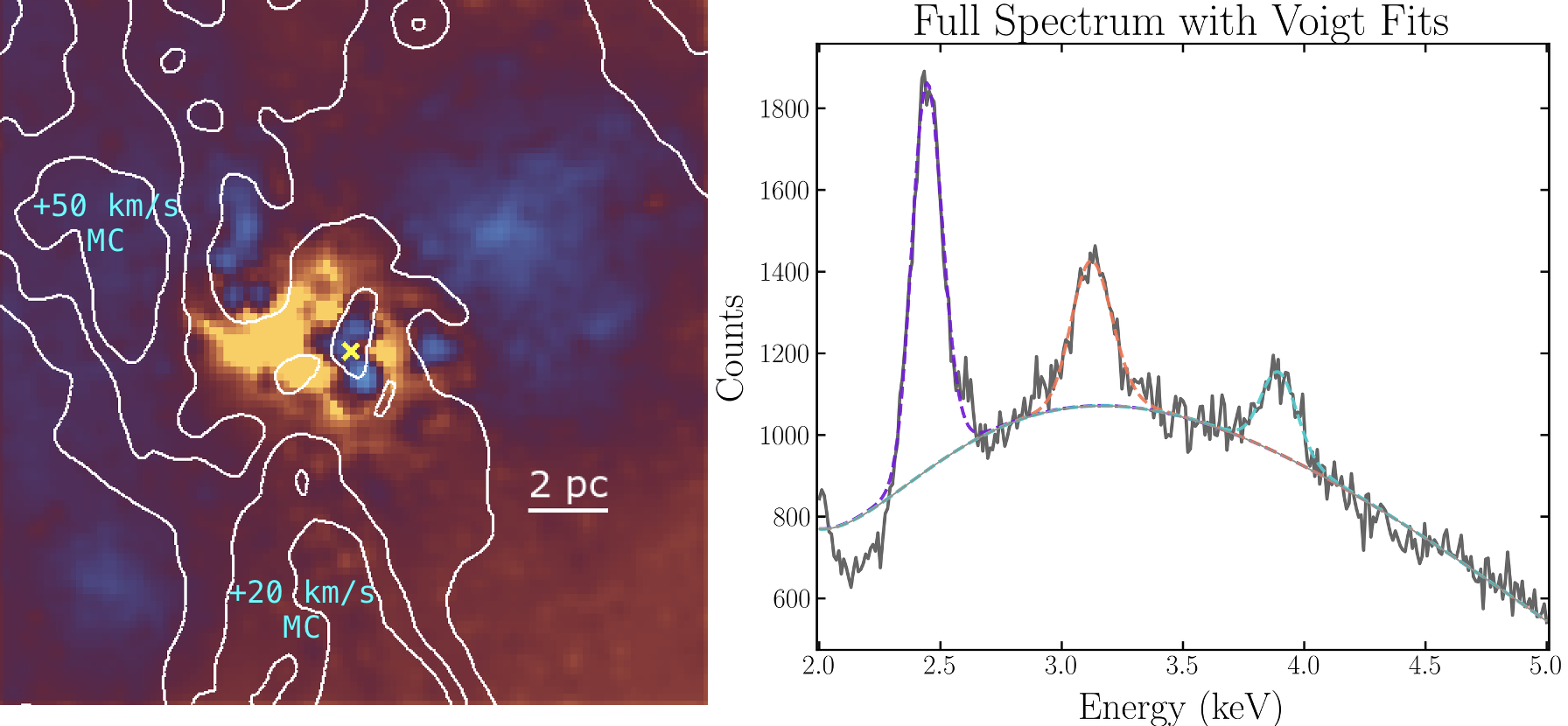}
    \caption{GMCA 2$-$5 keV Component 2. On the left, the image is overlaid with contours from the SCUBA 450 \micron\ continuum \citep{Pierce-Price2000}. The colorbar is consistent with the other pGMCA outputs, an ``X" marks the location of \SgrA, and a 2 pc scale bar is shown on the bottom right. On the right, the spectrum is dominated by a strong feature at $\sim$2.4 keV. This component exhibits the broadest spectral features of all pGMCA outputs, consistent with a composite spectrum formed by the superposition of many unresolved sources rather than emission from a single coherent plasma.}
    \label{fig:gmca_diffuse}
\end{figure*}

In Figure \ref{fig:gmca_diffuse}, we show the image and spectrum of the 2$-$5 keV Component 2. This spectral component is the most prominent among all pGMCA outputs, with the highest number of counts. Table~2 lists the measured spectral features with tentative identifications. The relative line strengths indicate a comparatively cool characteristic temperature of order $\sim$1~keV, as inferred from the weakness of higher-ionization features such as Ca~\textsc{xix} near 3.9~keV.

The spectral features show consistent broadening of $\Delta E \sim 0.2$--$0.25$~keV, the widest among the extracted components. Rather than indicating intrinsic turbulent line broadening in a single diffuse plasma, this behavior is naturally expected for a composite spectrum arising from the superposition of many unresolved stellar sources, primarily cataclysmic variables and active binaries, which together dominate the Galactic Center X-ray emission (GCXE) at these energies. In this sense, the CIE plasma models often used to describe this component provide a phenomenological representation of the underlying population-averaged emission rather than a unique physical plasma state.

Morphologically, this component is largely absent in the region coincident with the supernova remnant, but appears suppressed toward areas associated with the 20~$\mathrm{km \, s^{-1}}$ molecular cloud. It is brightest toward the 50~$\mathrm{km \, s^{-1}}$ cloud, the NE plume, material within the CND, the Western protrusion, and the emission features extending perpendicular to the Galactic plane that have been linked to past outflow activity \citep{Maeda2002} (see Figure~\ref{fig:ehm} for nomenclature). Its morphology shows minimal correlation with the dust emission maps used in this work, spanning dust temperatures from 10 to 1000~K, but does exhibit enhanced surface brightness in molecular gas cavities. This behavior is consistent with absorption and geometric effects modulating the observed emission from a broadly distributed GCXE component, rather than tracing localized heating or shock interaction.

The spatial distribution implies a line-of-sight geometry in which a substantial fraction of the GCXE lies between the observer and the 50~$\mathrm{km \, s^{-1}}$ material, while the 20~$\mathrm{km \, s^{-1}}$ cloud is located closer to the observer, more fully in the foreground. This configuration is consistent with previous kinematic and geometric studies of the region \citep[e.g.,][]{Coil2000}. The elevated emission observed in the NE plume, Western protrusion, and vertical features perpendicular to the Galactic plane likely reflects regions where molecular cavities reduce absorption, allowing emission from both the near and far sides of the GCXE to be observed. Meanwhile, the enhanced emission within the CND most plausibly reflects the high density of unresolved stellar sources and reduced effective absorption in the nuclear region, rather than emission from the wind-fed plasma associated with \SgrA, which is isolated separately by pGMCA.

\section{Using pGMCA Constraints for Spectral Fitting} \label{sec:spectral_fits}

In the previous sections, we applied pGMCA to \Chandra\ observations, which enabled us to separate spatially overlapping spectral components. Direct spectral fitting of the separated pGMCA spectra is not physically meaningful; instead, pGMCA provides physically motivated constraints and identifies regions of minimal cross-contamination that guide and improve spectral extraction in the Galactic Center.

\subsection{Extracting Regions}
Previous spectral analyses of the central parsec have consistently required multiple emission components. \citet{Ono2019} modeled the Suzaku Sgr A East spectrum with two CIE/\texttt{apec} and two recombining-plasma components. \citet{Maeda2002} used an \texttt{apec} + four Gaussian model to capture the prominent X-ray lines in the \Chandra\ spectrum of Sgr A East. Modeling the Galactic Center X-ray emission (GCXE) typically requires several thermal components to account for the strong Si/S, Ar, and Ca lines and the broad Fe complex (see, e.g., \citealt{Ono2019,Nobukawa2010}). Within $\sim5''$ of \SgrA, a RIAF-based model or hydrodynamic simulations are required to reproduce the accretion-flow emission \citep{Balakrishnan2024a,Balakrishnan2024b}. Even farther out, spectra of the wind-fed plasma around \SgrA\ generally require multi-temperature/shocked plasma descriptions to reach statistically acceptable fits. Most recently, XRISM modeling of Sgr A East \citep{XRISMCollaboration2024} required two broadened recombining-plasma components. These studies demonstrate that no single plasma process dominates anywhere within $\sim1'$ of \SgrA.

\begin{figure*}
    \centering
    \includegraphics[width=0.95\textwidth]{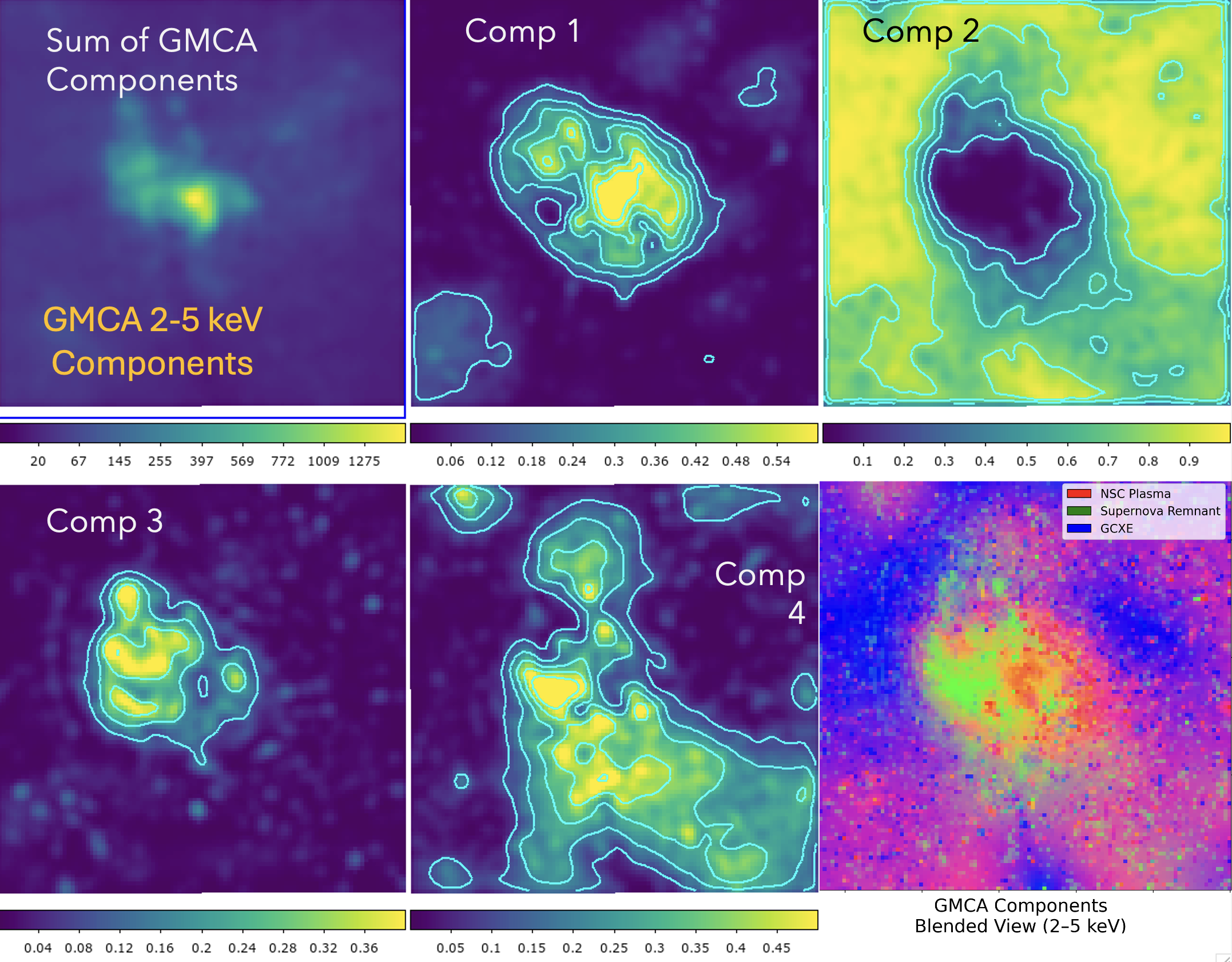}
    \caption{
        Fractional contributions of the four pGMCA components to the 2--5~keV emission. 
        \textit{Top left:} Sum of all pGMCA components. 
        \textit{Top middle/right; bottom left/middle:} Fractional maps for Components~1--4, respectively; cyan contours trace intensity structure.
        \textit{Bottom right:} Blended representation in which hue encodes component identity (green: supernova remnant; blue: GCXE; red: NSC/wind-fed plasma) and saturation reflects the fractional dominance at each pixel. 
        The spatial mixture of components underscores the difficulty of isolating clean spectra through conventional region selection alone.
    }
    \label{fig:gmca_fractional}
\end{figure*}

We calculated the fractional contribution of each pGMCA component to the observed emission. The fractional maps for Components~1--4 are shown in Figure~\ref{fig:gmca_fractional}. The top-left panel displays the total emission reconstructed from the sum of pGMCA components, while the bottom-right panel blends the components using a hue--saturation color mapping, clearly revealing the strong spatial mixing of distinct emission components in this region and illustrating why traditional extraction approaches struggle to isolate individual extended sources.

To mitigate pixel-scale noise, each fractional map was smoothed with a Gaussian kernel of width 1.5~pixels. For each pixel, we identified the dominant component (highest fractional value) and computed its ``gap,'' defined as the difference between the highest and second-highest fractional contributions. Large gaps mark pixels strongly dominated by a single component, while small gaps mark regions of mixed or uncertain origin. We adopted this adaptive dominance metric rather than applying fixed fractional thresholds (e.g., $>80\%$ contribution) to ensure stable relative emission from subdominant components in the resulting spectra.

\begin{figure}
    \centering
    \includegraphics[width=0.45\textwidth]{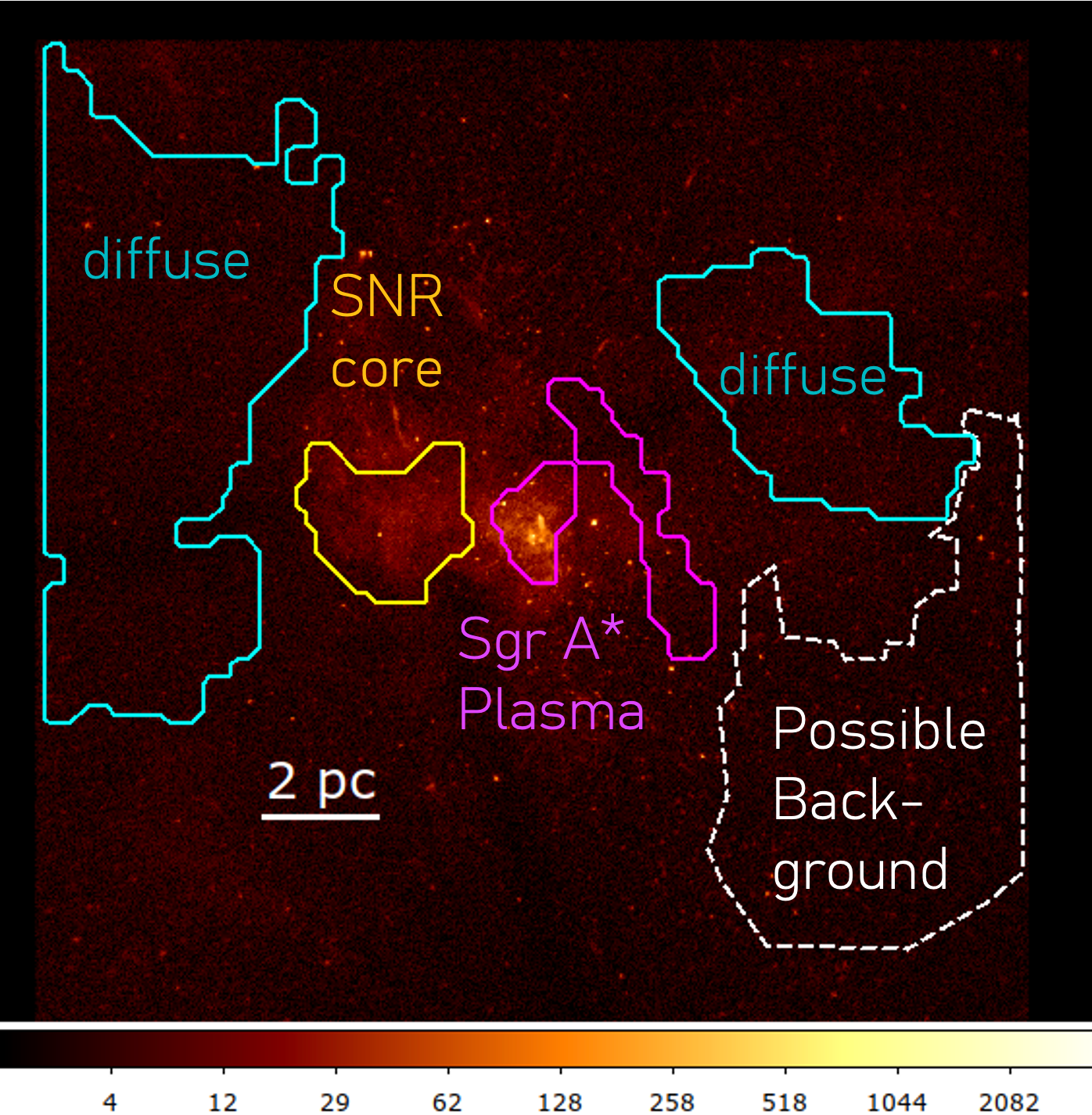}
    \caption{Spatially distinct spectral extraction regions derived from the pGMCA fractional maps, demonstrating regions where each component dominates. Contours mark the \textit{core} zones of each pGMCA component (see text). Yellow: Fe-rich Sgr A East core (SNR-dominated). Magenta: hot, wind-shocked plasma associated with \SgrA* (NSC/dust component), including a particularly bright western feature that may trace the interaction front between the Wolf--Rayet--driven outflow and the Western Arc/CND material located behind \SgrA\ in projection. Cyan: GCXE-dominated emission. The dashed white contour denotes a candidate background region in the Galactic plane that is free of bright structures and would, based on flux considerations alone, appear well suited for background subtraction within this field of view. However, as shown in Figure~\ref{fig:gmca_fractional}, this region in fact contains multiple emission components, and subtracting it would therefore not yield a cleanly background-subtracted spectrum. All regions are shown over the 2--8~keV \Chandra\ image. These regions enable pGMCA-informed spectral decomposition that minimizes cross-contamination between physically distinct emission components in the Galactic Center. Point sources were masked before spectral extraction.}
    \label{fig:spectralregions}
\end{figure}

We excluded pixels below the 30th percentile in total brightness to maintain sufficient signal-to-noise. Pixels in the top third of each component’s gap distribution were assigned to that component’s \textit{core} region, ensuring a consistent and physically motivated separation across the field. Only spatially contiguous structures larger than $\sim50$ pixels were retained to remove spurious islands. We then converted the cleaned masks into polygonal contours using the \texttt{shapely} library in world coordinates (FK5), producing DS9 region files for each component and for both core and transition zones. These extraction regions are shown in Figure~\ref{fig:spectralregions}. The fractional maps also highlight a bright feature dominated by the \SgrA\ wind--shocked plasma west of the black hole, which may represent the interaction front between Wolf--Rayet outflows and the Western Arc/CND material lying behind \SgrA\ in projection. This pGMCA-informed framework enables spatially resolved spectral extraction of physically distinct sources, allowing us to isolate, for example, the intrinsic SNR emission with minimized contamination from surrounding components.

\subsection{Spectral Modeling of the Sgr A East Core}

Using the pGMCA fractional maps, we calculated for each extraction region the relative contributions from three physical emission components: the NSC/wind-fed plasma (Component 1), the Sgr A East supernova remnant (Components 3 \& 4), and the GCXE (Component 2). These fractions were computed using counts-weighting (i.e., pixels with high photon counts had a greater influence on the mean values), to ensure that they trace each component’s true spectral contribution. We extracted spectra using standard \texttt{ciao} threads and masked point sources (see Appendix~\ref{appendix_sec:stacking}). The spectral regions where each component is dominant are shown in Figure~\ref{fig:spectralregions}. The low-surface-brightness ``background'' region indicated in Figure~\ref{fig:spectralregions} would appear suitable based on flux considerations alone; however, Figure~\ref{fig:gmca_fractional} shows that it contains substantial contributions from the \SgrA\ environment, the SNR, and the GCXE, and is therefore not an uncontaminated background.

In principle, the spectra can be described by:
\begin{equation}
Y_i(E) = \Sigma C_{ij} P_j(E)
\label{equation:sys}
\end{equation}
where $Y_i(E)$ is the observed spectrum of region $i$, $P_j(E)$ is the spectrum of emission component $j$, and $C_{ij}$ are the pGMCA-derived mixing coefficients. Table~\ref{tab:gmca_Aij} lists the coefficients $C_{ij}$ for each extraction region, where $j=1$ refers to the NSC/wind-fed plasma component, $j=2$ refers to the SNR plasma, and $j=3$ refers to the GCXE. In practice, jointly fitting multiple complex plasma models with tied parameters is computationally expensive and prone to local minima given current photon statistics. We therefore focus on the Sgr A East core spectrum, where pGMCA indicates a strong ($\gtrsim80\%$) dominance by the remnant emission.

\begin{deluxetable*}{rllll}
\centering
\tablewidth{0pt}
\tablecaption{Fractional Coefficients Returned by pGMCA for Different Regions.}
\tablehead{
\colhead{Region} & \colhead{$i$} & \colhead{C\(_{i1}\) (NSC/Dust)} & \colhead{C\(_{i2}\) (SNR core)} & \colhead{C\(_{i3}\) (GCXE)}
}
\startdata
Sgr A* Plasma & 1 & 0.7314 & 0.2139 & 0.0547 \\
SNR Core                 & 2  & 0.1467 & 0.8227 & 0.0306 \\
GCXE            & 3  & 0.0720 & 0.0186 & 0.9094 \\
\enddata
\tablecomments{Counts--weighted pGMCA mixing coefficients giving the fractional contribution of the NSC/wind-fed plasma (Component~1), SNR (Components~2+3), and GCXE (Component~4) to each extraction region. These coefficients, in combination with Equation~\ref{equation:sys}, define a pGMCA-informed joint spectral decomposition framework.}
\label{tab:gmca_Aij}
\end{deluxetable*}

Previous X-ray studies have demonstrated that Sgr A East requires a multi-component spectral model (combinations of \texttt{apec}, \texttt{nei}, and \texttt{gauss} components) to reproduce its rich line emission (e.g., \citealt{Maeda2002,Ono2019,XRISMCollaboration2024}). We did not subtract a local, source-free background region for this spectrum; instead, we utilized ACIS blank-sky files extracted and normalized using standard methods. Additionally, the low-emission ``background'' region that might typically be chosen within the field (Figure~\ref{fig:spectralregions}) actually contains substantial spectral contributions from the GCXE and the NSC/wind-fed plasma, reinforcing the need to explicitly model these components.

We explored a range of thermal plasma descriptions in XSPEC, informed by the pGMCA mixing analysis. The \texttt{vnei} component was adopted to represent Fe-rich ejecta, allowing Si, S, Ar, Ca, and Fe abundances to vary between $0.5$--$5\,Z_\odot$, while $kT$ was permitted to vary between $0.5$ and $10$~keV. The ionization age $\tau$ was constrained to span $10^{9}$--$10^{12}\,\rm cm^{3}\,s^{-1}$. For the plasma associated with \SgrA, we found that a multi-temperature \texttt{cevmkl} model provided a significantly better fit than single-plasma alternatives such as \texttt{vpshock} or \texttt{vnei}, consistent with the expectation that the wind-shocked flow near \SgrA\ has a complex temperature and velocity structure \citep{Balakrishnan2024b}. For the GCXE contribution, we used a \texttt{vapec + gauss} combination with abundances fixed to $2\,Z_\odot$ following previous work \citep[e.g.,][]{Ono2019}; this term is used as a phenomenological description of residual GCXE spectral structure along the line of sight, which is widely interpreted as being dominated by unresolved stellar sources at these energies. Overall, the following configuration was found to be the simplest model capable of capturing the combined spectral signatures of the SNR, the \SgrA\ wind-fed plasma, and residual GCXE emission:

\begin{center}
\texttt{tbabs*const*vnei + tbabs*const*cevmkl + tbabs*const*(vapec+gauss)} 
\end{center}

This model produced a good fit to the blank-sky subtracted spectrum with $\chi^2 /N_{\rm bins} = 55/47 = 1.17$. When individual components were removed, the fit degraded: excluding the GCXE term increased $\chi^2$ to $301/47$, and excluding the \SgrA\ wind-fed plasma increased $\chi^2$ to $613/47$. Removing the pGMCA-derived coefficients entirely (while retaining all three components) resulted in a fit statistic of $262/47$. Applying a model that has provided statistically good fits on background-subtracted versions of the remnant spectrum, \texttt{tbabs*(vnei+vnei)}, gave a fit statistic of $230/47$. In summary, although the non-SNR components contribute a smaller fraction of the total flux in this region, their distinct line and continuum features must be included to accurately model the Sgr A East core spectrum.

\begin{deluxetable*}{rlll}
\tablewidth{0pt}
\tablecaption{Best-fit plasma parameters and 90\% confidence intervals for the pGMCA-informed Sgr A East spectral region.}
\tablehead{
\colhead{Parameter} & \colhead{SNR Comp} & \colhead{\SgrA\ Plasma Comp} & \colhead{GCXE Comp}
}
\startdata
 N$_H$ ($10^{22} \rm cm^{-2})$& 15.4$\substack{+3.1\\-1.7}$& 19.8 $\substack{+0.2\\-0.4}$& 11.6 $\substack{+0.7\\-1.2}$\\
     $kT$ (keV)&  4.1 $\substack{+0.5\\-1.7}$ &  1.3 $\substack{+0.04\\-0.02}$ & 0.9 $\pm~0.1$\\
     Si (Z$_\odot$) &  4.95$\substack{+0.05\\-0.5}$ &  4.2 $\pm~0.2$ & 2 (fixed)\\
     S (Z$_\odot$) &  0.75 $\substack{+0.4\\-0.2}$ &  1.2 $\substack{+0.1\\-0.2}$ & 2 (fixed)\\
     Ar (Z$_\odot$) &  0.7 $\substack{+0.1\\-0.2}$  &  1.8 $\substack{+0.1\\-0.3}$ & 2 (fixed)\\
     Ca (Z$_\odot$) &  4.9 $\substack{+0.1\\-1.0}$ &  2.1 $\substack{+0.3\\-0.5}$ & 2 (fixed)\\
     Fe (Z$_\odot$) &  4.9 $\substack{+0.1\\-0.8}$ &  1.9 $\substack{+0.1\\-0.4}$  & 2 (fixed)\\
     Ionization Age (cm$^{3}$/s)&  3.7 $\substack{+1.3 \\ -0.6} \times 10^{9}$&  -& -\\ 
     Gaussian $E_0$ (keV) & - & - & 6.643 $\pm~0.007$ \\
     Gaussian $\sigma$ (keV) & - & - & 0.05 $\pm~0.01$  \\ 
\enddata
\tablecomments{The best-fit spectral model for the Sgr A East region yields a reduced chi-square of $\chi^2/\nu = 55/47 = 1.17$. The metallicities were constrained to vary between $0.5\,Z_\odot$ and $5\,Z_\odot$. The supernova remnant plasma is modeled with \texttt{tbabs*vnei}, while the surrounding \SgrA\ environment is modeled with \texttt{tbabs*cevmkl} (here, $kT$ corresponds to the maximum temperature of the multi-temperature distribution, \texttt{T$_{max}$}). For the GCXE term, modeled as \texttt{tbabs(vapec + gauss)}, we fixed the abundance to $2\,Z_\odot$ following previous studies \citep{Ono2019}. By separating these spectral components, we infer a lower ionization age for the Sgr~A~East plasma than reported in most previous analyses, with $\tau = 3.7 \times 10^{9}\mathrm{cm^{-3}/s}$.}
\label{tab:modelfit}
\end{deluxetable*}

Table~\ref{tab:modelfit} presents the best-fit spectral parameters for the three components required to reproduce the Sgr~A East spectrum: the Fe-rich supernova ejecta, the \SgrA\ wind-fed plasma, and the GCXE contribution along the line of sight. The \texttt{const} values in the model are taken from Table~\ref{tab:gmca_Aij}. We tried allowing these mixing coefficients to vary up to 0.1, but this did not increase the goodness-of-fit. The absorbing column densities differ substantially among components, with the \SgrA\ plasma exhibiting the highest $N_H$, consistent with its partial association with dense dust structures along the line of sight \citep{Lau2015}, while the GCXE are subject to lowest effective column. The SNR ejecta are characterized by a high temperature ($kT \sim 4$~keV), strongly supersolar Si, Ca, and Fe abundances, and a low ionization age of $\tau = (3.7^{+1.3}_{-0.6}) \times 10^{9}\,\rm cm^{3}\,s^{-1}$, indicating metal-rich plasma that remains far from collisional ionization equilibrium. In contrast, S and Ar are not strongly enhanced. Models forced to $\tau \gtrsim 10^{11},\rm cm^{3},s^{-1}$ do not produce the observed Fe–K line centroid without unphysical abundances to compensate. High Si, Ca, and Fe but comparatively weak S and Ar is qualitatively consistent with emission dominated by inner ejecta layers associated with incomplete and complete Si burning, rather than fully mixed core-collapse supernova yields, as seen in modern nucleosynthesis models \citep{WoosleyWeaver1995,Sukhbold2016,LimongiChieffi2018}. In these models, S and Ar originate primarily in explosive O-burning layers and are therefore more sensitive to progenitor structure, explosion energy, and mixing than Si, Ca, and Fe, such that incomplete mixing or selective sampling of inner ejecta can naturally suppress their apparent abundances \citep{Thielmann1996,Nomoto2006}. The \SgrA\ plasma is significantly cooler ($kT \sim 1.3$~keV), moderately supersolar in $\alpha$-elements, and does not require a non-equilibrium ionization treatment, consistent with a continuously shocked, multi-temperature flow that is better described by a \texttt{cevmkl} model. The GCXE term exhibits the lowest characteristic temperature and includes a broad Fe~K feature centered at $6.64$~keV, consistent with He-like Fe emission.

\section{Discussion} \label{sec:summary}

In this work, we have presented the first application of pGMCA on \Chandra\ observations of the central $\sim$30~pc. Rather than relying on band-limited images that inevitably mix line-of-sight components, pGMCA allows physically interpretable plasmas to be isolated in both image and spectral space. These components were hinted at through narrow-band imaging in previous studies \citep[see, e.g.][]{Maeda2002, Wang2022}, but here they are recovered as spatially coherent, spectrally distinct structures that can be traced across the field without assuming a priori region boundaries. We demonstrate the power of using signal-separation techniques to better understand complex, interacting environments; using the pGMCA images together with multiwavelength observations, we constrain the geometry and interaction sites between hot plasma and cold material, and the spatially resolved spectral information provides direct, physically motivated priors for robust spectral fitting in a region where traditional background selection is intrinsically contaminated.

\subsection{Multiwavelength Comparisons \& Morphological Interpretations}

In the 2--5~keV band, pGMCA separated two morphologically distinct Sgr~A East-associated structures that are both centered on the radio shell (middle and right panel of Figure~\ref{fig:gmca_snr}), supporting their association with the remnant. 2--5~keV Component~3 has strong lines from S~\textsc{xv}, Ar~\textsc{xvii}, and Ca~\textsc{xix}, consistent with a multi-temperature, chemically stratified plasma that traces ejecta mixed with swept-up material. Comparing these results to median-energy maps, we see that the Fe-rich core is co-spatial with regions of high median photon energy, while the S/Ar/Ca-enhanced component extends farther outward, encompassing areas with softer emission and mapping a more extended, lower-ionization/temperature envelope around the core.
    
GMCA 2--5~keV Component~3 is clearly interacting with the molecular filament. Regions of low median photon energy lie where dense molecular gas is absent, indicating that these sightlines suffer minimal soft-photon absorption. Meanwhile, the central regions of the remnant appear harder in projection due to molecular material in the foreground that has not been vaporized. Comparing to 25.7~K dust maps, we see a sharp interface between the pGMCA morphology and the dust emission inside which compressed ejecta has thermalized. Moreover, maser emission at both 0~$\mathrm{km\,s^{-1}}$ and 50~$\mathrm{km\,s^{-1}}$ along this boundary confirms that the shock is expanding outward from within the molecular cloud in both directions along our line of sight.

Previous SOFIA/FORCAST observations of warm dust highlighted by \citet{Lau2015} revealed a compact region within Sgr~A East at an average temperature of $\sim$100~K, indicating a population dominated by very small grains ($\sim$0.01~\micron). The co-location of our pGMCA morphology with the \citet{Lau2015} dust feature, along with the lack of X-ray emission from nearby H\,\textsc{ii} regions, supports the interpretation that the warm dust is heated locally by SNR-driven shocks (rather than by photoionization from adjacent H\,\textsc{ii} regions) and that it has endured the reflected shock while remaining embedded within the remnant interior.

GMCA 2--5~keV Component~2 appears to be associated with extended hot plasma. It is anti-correlated with the supernova remnant but shows spatial overlap with several notable structures, including the NE plume, material within the CND, the Western protrusion, soft X-ray features previously linked to past outflows from \SgrA\ \citep{Maeda2002}, and the 50~$\mathrm{km\,s^{-1}}$ component of the molecular filament (see Figure~\ref{fig:ehm} for labelling). Its apparent absence in regions associated with the 20~$\mathrm{km\,s^{-1}}$ cloud likely results from line-of-sight absorption, implying that the 20~$\mathrm{km\,s^{-1}}$ material lies between us and the majority of the plasma. The plasma within the CND is likely bright due to the presence of abundant optically thin, turbulent plasma generated by the stellar winds. In contrast, the enhanced emission seen in the NE plume, Western protrusion, and ejections perpendicular to the Galactic plane likely arises from molecular cavities in those regions, allowing us to see the near and far side of the plasma.

\subsection{Isolating the SNR in Spectral Modeling}

Our spectral modeling of the Sgr~A East core uses pGMCA-imposed imaging constraints to implement a three-component plasma model consisting of (i) a hot, Fe-rich \texttt{vnei} ejecta component, (ii) a cooler, multi-temperature \texttt{cevmkl} component tracing the \SgrA\ wind-shocked plasma, and (iii) a \texttt{vapec + gauss} component representing residual GCXE emission. As mentioned previously, we utilize the ACIS-I spectrum as it has a lower background and increasing the quality of certain parts of our analyses. However, ACIS-S has detected Cr, Mn, and Ni lines associated with the remnant, and is able to distinguish between the 6.13 Mn K$\alpha$, 6.6 Fe K$\alpha$, and the 6.9 Fe Lyman$\alpha$ lines. Therefore, a prudent analysis would be to apply this framework to the ACIS-S Chandra data, but that is beyond the scope of this work.  We find that a fit to the ACIS-I data requires both high-temperature, Fe- and Ca-enhanced ejecta and a cooler plasma with comparatively weak S and Ar, a pattern that is qualitatively consistent with emission dominated by inner core-collapse supernova ejecta layers rather than fully mixed nucleosynthetic yields, as seen in both classic and modern yield calculations \citep{WoosleyWeaver1995,Nomoto2006,Sukhbold2016,LimongiChieffi2018}. Our most notable quantitative result is the low ionization age inferred for the ejecta, $\tau = (3.7^{+1.3}_{-0.6}) \times 10^{9}\,\rm cm^{3}\,s^{-1}$, nearly two orders of magnitude smaller than values inferred from spatially integrated NEI analyses of Sgr~A East. Previous \Chandra\ and \textit{XMM-Newton} studies generally find $n_e t \gtrsim 10^{12}\,\rm cm^{3}\,s^{-1}$, consistent with plasma near collisional ionization equilibrium \citep{Sakano2004,Park2004}, while recent XRISM/Resolve NEI modeling yields $\tau \sim 10^{11}\,\rm cm^{3}\,s^{-1}$ for the Fe-rich plasma \citep{XRISMCollaboration2024}. For plausible ejecta densities of $n_e \sim 1$--$10\,\rm cm^{-3}$, our measured $\tau$ would correspond to a post-shock timescale of only $\sim10$--$10^{2}$~yr. However, our results show clear evidence for shock compression associated with interaction with the $50\ \rm km\,s^{-1}$ molecular cloud, implying substantially higher post-shock electron densities. In this case, a denser plasma with $n_e \sim 10^{2}$--$10^{4}\,\rm cm^{-3}$ and a correspondingly low ionization temperature provides a more physically plausible description of the shocked ejecta. Alternatively, part of the discrepancy may arise from modeling degeneracies, including tradeoffs between temperature structure and ionization age or sensitivity to Fe--K line modeling in the ejecta-dominated spectrum.

\section{Summary \& Conclusions}

Our analysis began by creating diffuse emission maps and median-energy maps to identify key features in the X-ray data, which provide the foundation for interpreting the subsequent signal-separation algorithm derived results. We then applied the method pGMCA to 1.3~Ms of stacked, point-source-removed ACIS-I observations and compared the recovered components to complementary multiwavelength datasets to establish physically motivated interpretations. The algorithm is particularly effective at isolating line-dominated emission from the supernova remnant (Figure~\ref{fig:gmca_snr}, left panel), whose center is clearly offset from broader Fe-dominated emission in the 5--8~keV band (Figure~\ref{fig:gmca_dust}), which instead traces the central stellar cluster and circumnuclear disk (CND).

Overall, pGMCA isolates regions dominated by the NSC/wind-fed plasma, SNR emission, and the GCXE (Figure~\ref{fig:gmca_fractional}). The NSC/wind-fed plasma is concentrated toward the central region and traces the extended, shock-heated stellar-wind medium around \SgrA, including a bright western enhancement that plausibly marks an interaction front with the Western Arc/CND material. The GCXE contributes a comparatively smooth, diffuse baseline across much of the field of view. In contrast, the SNR morphology cleanly recovers the Fe-rich core and surrounding shell of Sgr~A East. The pGMCA-derived mixing coefficients (Table~\ref{tab:gmca_Aij}) show that the Sgr~A East core aperture is dominated ($\gtrsim80\%$) by SNR emission, with smaller but non-negligible contributions from the NSC/wind-fed plasma and the GCXE.

These results demonstrate that pGMCA-informed modeling provides a self-consistent pathway for separating the spectral signatures of multiple plasma structures in the innermost Galactic Center and for isolating the intrinsic emission of Sgr~A East despite contamination from the \SgrA\ wind-shocked plasma and diffuse GCXE emission. The results reinforce the picture of Sgr~A East as a mixed-morphology SNR evolving in a dense, structured environment, now resolved into a Fe-rich core and surrounding S/Ar/Ca-enhanced plasma. The remnant has expanded into, compressed, and shock-heated the surrounding molecular material, likely carving cavities that shape the three-dimensional structure of the region. 

The remnant’s proximity to \SgrA\ suggests a mutual influence, in which Sgr~A East has both impacted and been shaped by the evolution of the central SMBH environment. The SNR spectrum indicates temperatures of $kT \sim 2$--$5$~keV out to $\sim2$~pc \citep{baganoff2003}, implying inefficient radiative cooling: for plasma at $n_e \sim 1$~cm$^{-3}$ and $kT \sim 3$~keV, the cooling time is $t_{\rm cool} \sim 10^{7}$--$10^{8}$~yr \citep{Draine2011}, far longer than the estimated SNR age of $\sim10^{4}$~yr. The persistent contribution of both the \SgrA\ wind-fed plasma and the GCXE further shows that Sgr~A East does not evolve in isolation but remains energetically and spectrally entangled with the broader Galactic Center ecosystem. At the same time, shock compression traced by the pGMCA 2--5~keV SNR component and the 25.7~K dust filament indicates that the remnant may promote localized gas densification while suppressing fragmentation within the shell. Shock-excited OH masers and surviving warm dust further attest to the remnant’s ongoing interaction with dense gas \citep{YZ1997}. Together, these results identify Sgr~A East as a major agent of energy injection that has kept the circumnuclear environment of \SgrA\ hot and dynamically turbulent over its $\sim10^{4}$~yr lifetime.

\section{Software and third party data repository citations} \label{sec:cite}

\vspace{5mm}
\facilities{Chandra (ACIS), SCUBA (James Clark Maxwell Telescope), Herschel, VLA, SOFIA FORCAST, HST NICMOS, ALMA, VLA}


\software{Astropy \citep{astropy:2013, astropy:2018, astropy:2022}, pGMCA \citep{Bobin2020}, Matplotlib \citep{Hunter2007}
          }


\begin{acknowledgments}
MB thanks Giovanni Stel for providing the HCN contours and Dylan Pare for providing the SOFIA HAWC+ emission and magnetic field lines. MB also thanks Masato Tsuboi for putting all his cleaned images on the ALMA archive for easy access. QDW acknowledges support from SAO/CXC/NASA through grants GO3-24120X and GO4-25091X.
\end{acknowledgments}

\appendix

\section{Stacking the Chandra ACIS-I Observations} \label{appendix_sec:stacking}

We began by retrieving all available Chandra ACIS-I observations of the Galactic Center with exposure times exceeding 10 ks (listed in Table \ref{tab:chandra_data}) from the Chandra Data Archive, as exposures often did not have enough counts to be used by \texttt{celldetect} to align observations. Standard data processing was performed using the CIAO \texttt{chandra\_repro} tool. To filter out background flares, we applied the \texttt{deflare} script, selecting time bins based on visual inspection of light curves. We restricted the light curve energy range to 2~keV, below the threshold where Galactic absorption ($N_H \sim 10^{23} \ \rm cm^{-2}$)  renders the source flux negligible, identifying only particle-induced flares, not X-ray flares from \SgrA. Since our analysis focuses on arcminute-scale diffuse structure, the presence of flares from \SgrA, which manifest on arcsecond scales, does not significantly affect our results. We prioritized retaining as much of the large-scale signal as possible.

\begin{figure}
    \centering
    \includegraphics[width=0.58\textwidth]{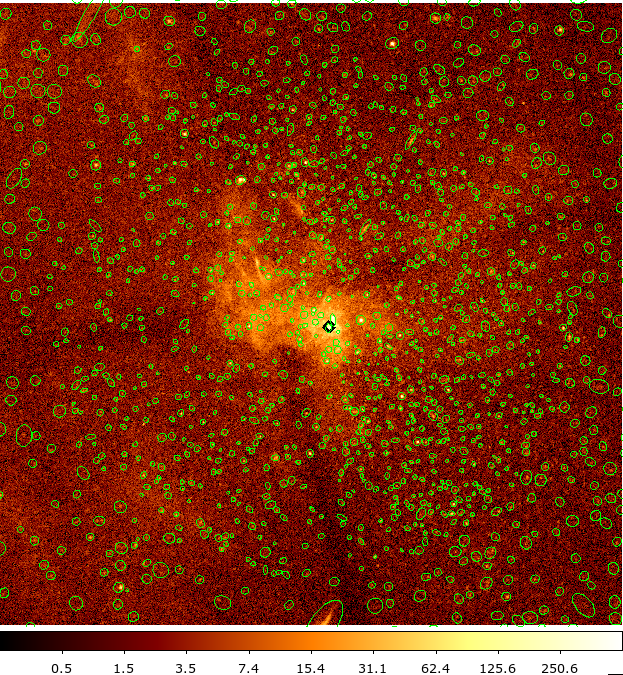}
    \caption{Merged \Chandra\ event image created using \texttt{reproject\_obs}, with point sources identified by \texttt{wavdetect} shown as green ellipses. These regions were manually refined to exclude diffuse structures misidentified as compact sources. The black diamond near the center marks the location of the supermassive black hole, \SgrA. The image is centered on \SgrA\ and spans 8.4' x 8.4'. }
    \label{fig:sources}
\end{figure}

Astrometric alignment was performed using the CIAO tools \texttt{celldetect} and \texttt{wcs\_match}, aligning all observations to the tangent point of OBSID 3392 following the standard aspect re-projection procedure. For preliminary alignment and construction of the pGMCA input cubes, we used source lists generated with \texttt{celldetect}, which is substantially faster than \texttt{wavdetect} and sufficient for identifying bright sources in this crowded field. We experimented with several detection configurations and ultimately adopted \texttt{fixedcell=6} and \texttt{thresh=6}, using a fixed cell size to avoid biases introduced by incomplete exposure maps. The resulting source list was manually refined: spurious detections were removed, and overlapping detections associated with filamentary structures were merged into single extended elliptical regions. Using the aligned event files, we then produced exposure-corrected images in multiple energy bands (1–2.6 keV, 2.6–5 keV, 4.5–6 keV, 5–8 keV, and 2–8 keV) with \texttt{flux\_obs} and \texttt{merge\_obs}, weighting each exposure map by the effective exposure time. The final set of removed point sources is shown in Figure~\ref{fig:sources}, overlaid on the 1.3 Ms merged event file. 

To isolate diffuse X-ray structures, we identified point sources using the CIAO tool \texttt{wavdetect}, which correlates the image with a Mexican Hat wavelet function at multiple spatial scales. We adopted wavelet scales of 4, 8, and 16 pixels to ensure sensitivity to both compact and moderately extended sources. The significance threshold was set to produce approximately one spurious detection per CCD. At each scale, \texttt{wavdetect} performs an iterative cleansing procedure, masking out significant features to estimate the local background before detecting statistically significant peaks. We specified \texttt{ellsigma=2} to enlarge the resulting elliptical source regions, enabling conservative masking of source flux.

Following source detection, we manually reviewed and edited the \texttt{wavdetect} output. Several sources—particularly in regions of bright diffuse emission—appeared to be compact knots within extended structures rather than discrete point sources. These were removed from the region list to avoid over-subtraction. We verified that no obvious residual point sources were present in the resulting diffuse emission maps.

We chose not to exclude the filamentary structures previously identified as nonthermal filaments \citep[see, e.g.][]{YZ2005}. Their inclusion or exclusion has no effect on the GMCA results, as the individual filaments are not of comparable strength with the overall Sgr A East or Sgr~A* plasma emission. We note that while identifying and excluding point sources is important for accurate imaging of diffuse structures, our primary results, particularly those based on pGMCA decomposition, are not highly sensitive to the exact extent of source removal. pGMCA is robust against residual point sources, and our conclusions remain consistent across different masking strategies.

\bibliography{sample631}{}
\bibliographystyle{aasjournal}



\end{document}